\documentclass[%
 aip,
 amsmath,amssymb,
 reprint,%
]{revtex4-1}

\usepackage{graphicx}
\usepackage{dcolumn}
\usepackage{bm}

\usepackage[utf8]{inputenc}
\usepackage[T1]{fontenc}
\usepackage{mathptmx}
\usepackage{etoolbox}

\usepackage{booktabs}

\DeclareMathOperator*{\argmin}{arg\,min}

\makeatletter
\def\@email#1#2{%
 \endgroup
 \patchcmd{\titleblock@produce}
  {\frontmatter@RRAPformat}
  {\frontmatter@RRAPformat{\produce@RRAP{*#1\href{mailto:#2}{#2}}}\frontmatter@RRAPformat}
  {}{}
}%
\makeatother
\begin{document}

\preprint{AIP/123-QED}

\title[Grad-Shafranov equilibria via data-free physics informed neural networks]{Grad-Shafranov equilibria via data-free physics informed neural networks}
\author{Byoungchan Jang}
 \email{byoungj@umd.edu}
 \affiliation{Institute for Research in
Electronics and Applied Physics, University of Maryland, College Park, MD, 20742, USA\looseness=-1}
\author{Alan A. Kaptanoglu}%
 \affiliation{Courant Institute, New York University, New York, NY, 10012, USA\looseness=-1}
\author{Rahul Gaur}
 \affiliation{Department of Mechanical and Aerospace Engineering, Princeton University, Princeton, New Jersey 08544, USA\looseness=-1}
\author{Shaowu Pan}
\affiliation{%
 Department of Mechanical, Aerospace and Nuclear Engineering, Rensselaer Polytechnic Institute, Troy, NY, 12180, USA\looseness=-1
 }%
\author{Matt Landreman}
 \affiliation{Institute for Research in
Electronics and Applied Physics, University of Maryland, College Park, MD, 20742, USA\looseness=-1}
\author{William Dorland}
 \affiliation{Institute for Research in
Electronics and Applied Physics, University of Maryland, College Park, MD, 20742, USA\looseness=-1}

\date{\today}

\begin{abstract}
A large number of magnetohydrodynamic (MHD) equilibrium calculations are often required for uncertainty quantification, optimization, and real-time diagnostic information, making MHD equilibrium codes vital to the field of plasma physics. In this paper, we explore a method for solving the Grad-Shafranov equation by using Physics-Informed Neural Networks (PINNs). For PINNs, we optimize neural networks by directly minimizing the residual of the PDE as a loss function. We show that PINNs can accurately and effectively solve the Grad-Shafranov equation with several different boundary conditions. We also explore the parameter space by varying the size of the model, the learning rate, and boundary conditions to map various trade-offs such as between reconstruction error and computational speed. Additionally, we introduce a parameterized PINN framework, expanding the input space to include variables such as pressure, aspect ratio, elongation, and triangularity in order to handle a broader range of plasma scenarios within a single network. Parametrized PINNs could be used in future work to solve inverse problems such as shape optimization. 
\end{abstract}

\maketitle


\section{\label{sec:intro}Introduction}

The goal of magnetic confinement fusion research is to establish a hot, dense, steady-state plasma with enough fusion occurring to be economically viable. For fusion parameter regimes, the plasma is often well-described by ideal magnetohydrodynamics (MHD), and 2D equilibria are described by the Grad-Shafranov equation (GSE).~\cite{grad1958hydromagnetic,shafranov1958magnetohydrodynamical} For diagnostic and real-time control purposes on current quasi-steady-state fusion devices, accurate plasma equilibria must be computed in real-time from a sparse set of measurement data. Iterative solves of the full GSE can be expensive, and the choice of basis functions and boundary conditions for the pressure and current profiles is crucial.~\cite{lao2005mhd} These types of calculations can be improved and sped up with neural network surrogate models. Neural network surrogates provide flexibility since the solution is differentiable and meshless. Once these surrogate models are trained, the inference times are very fast and additional improvements in GPUs and neural networks can further improve the performance. Moreover, they can be useful for uncertainty quantification (UQ), as the evaluation of these surrogate models post-training is very inexpensive.

In a physics-informed neural network (PINN), the weights and biases of the network are chosen by directly minimizing
the residual of a partial differential equation (PDE) as a loss function.~\cite{raissi2017physics1} Accordingly, PINNs do not need training data obtained through potentially expensive physical experiments or computer simulations.~\cite{raissi2017physics2} PINNs can be used to solve many types of PDEs, including high-dimensional partial differential equations~\cite{blechschmidt2021three} and fractional partial differential equations.~\cite{karniadakis2021physics} PINNs can also be used to solve various inverse problems, such as shape optimization,~\cite{shukla2023deep,rossi2023potential} that derive from PDEs.~\cite{cai2021physics} In this paper, we use PINNs to solve the fixed-boundary Grad-Shafranov equation under two variations of the boundary conditions -- ``soft'' and ``hard'' -- and show that PINNs can accurately and effectively solve the equation.

More than two decades ago, neural networks were proposed and briefly studied for solving the ideal MHD force balance equation for tokamaks and stellarators.~\cite{van1995neural} However, recently there has there been a resurgence of interest in neural network surrogates for MHD equilibria.~\cite{joung2019deep,merlo2021proof,wai2022neural,liu2022surrogate} Most of this work has not used physics-informed neural networks (i.e. not directly minimizing the residual of the ideal MHD force balance) or has used substantial experimental data to build a fast surrogate for a specific device. There are also cases where the force balance residual is used as physics regularization~\cite{merlo2023physics} or the MHD force balance is solved with a pseudospectral method via a fully spectral polynomial-Fourier basis instead of a typical neural network 

\subsection{\label{sec:level2}Contributions of this work}

In the present work, we generate fully data-free and mesh-free surrogate models of the Grad-Shafranov equation via physics-informed neural networks (PINNs). The intent of this work is not to replace optimized GS solvers. We aim to extend the proof-of-concept PINNs from Kaltsas et al.~\cite{kaltsas2022neural} to future work such as device-agnostic GS solvers and 3D equilibrium solvers without nested flux assumptions. We illustrate the strength of this approach by showing solutions of a class of fixed-boundary ``one-size fits all'' tokamak configurations, including single and double null configurations.~\cite{cerfon2010one} Configurations are implemented with both soft and hard constraints to compare their performance. We also perform hyperparameter scans and other systematic investigations and conclude that the network produces robust results for a very broad range of network parameters. We conclude by demonstrating preliminary work on \textit{parametric} PINNs that are parametrized by the magnitude of the pressure and three shape parameters associated with the ``elongated-D'' cross section used for tokamak shapes. Unlike standard PINNs in which the models need to be retrained for each boundary shape and choice of profiles, parametric-PINNs only need to be trained once for various configurations. After a modest training period, a parametric-PINN can infer equilibria in tens of milliseconds for different geometries, from negative triangularity tokamaks to field-reversed configurations.
For the cases explored here, the solution accuracy is likely sufficient to use a parametric PINN as a surrogate in inverse problems and optimization, although we have not yet explored these applications. Implementing PINNs in this work was done through a modified version of DeepXDE, a Python library for physics-informed learning.~\cite{lu2021deepxde}

\section{\label{sec:MHD_Equilibria}MHD equilibria}
MHD equilibria are essential in understanding fusion devices such as tokamaks and stellarators. Equilibrium solutions are used for many applications including diagnostic interpretation, the application of real-time control, and the identification of instability. For toroidal or cylindrical geometries with a symmetry direction, the Grad-Shafranov equation (GSE) defines the equilibria. The GSE is derived from general ideal-MHD force-balance, 
\begin{align}
\label{eq:force_balance}
     J \times B = \nabla p.
\end{align}
Given a symmetry direction, the divergence-free condition,
\begin{align}
\label{eq:full_gse}
     \nabla\cdot B = 0,
\end{align}
 and some vector algebra, we can reduce the force balance equation to the GSE. Details of the derivation can be found in various textbooks.~\cite{freidberg2014ideal,hazeltine2003plasma} In short, for an axisymmetric configuration, a poloidal magnetic flux function $\psi$ is introduced where
\begin{align}
\label{eq:magnetic_flux_function}
    B &= \frac{1}{r} \nabla\psi\times e_{\phi} + \frac{F}{r}e_{\phi}\\
    \mu_0 J &= \frac{1}{r} \frac{dF}{d\psi}\nabla\psi\times e_{\phi} -      \frac{1}{r}\left[  r\frac{\partial}{\partial{r}} \left(\frac{1}{r}\frac{\partial{\psi}}{\partial{r}}\right) + \frac{\partial^2{\psi}}{\partial{z^2}} \right] e_{\phi}.
\end{align}
Here $\phi$ is the toroidal angle, $r$ is the major radial coordinate in cylindrical coordinates, $e_{\phi}$ is the unit vector in the toroidal direction, and $F(\psi) = rB_{\phi}$.

The GSE describes static 2D MHD equilibria via the poloidal flux function $\psi(r, z)$ in standard cylindrical coordinates $(r, \phi, z)$:
\begin{align}
\label{eq:full_gse}
    \psi_{rr} - \frac{1}{r}\psi_r + \psi_{zz} + \mu_0 r^2\frac{dp(\psi)}{d\psi} + \frac{1}{2}\frac{dF^2(\psi)}{d\psi} = 0.
\end{align}
Typically, the pressure profiles $p$ and poloidal current profiles $F = rB_\phi$ are free functions that must be prescribed to solve the PDE. In other words,
given the pressure and $F$ profiles, solving~\eqref{eq:full_gse} provides the MHD equilibrium.

\begin{figure}[b]
    \centering
    \includegraphics[width=\linewidth]{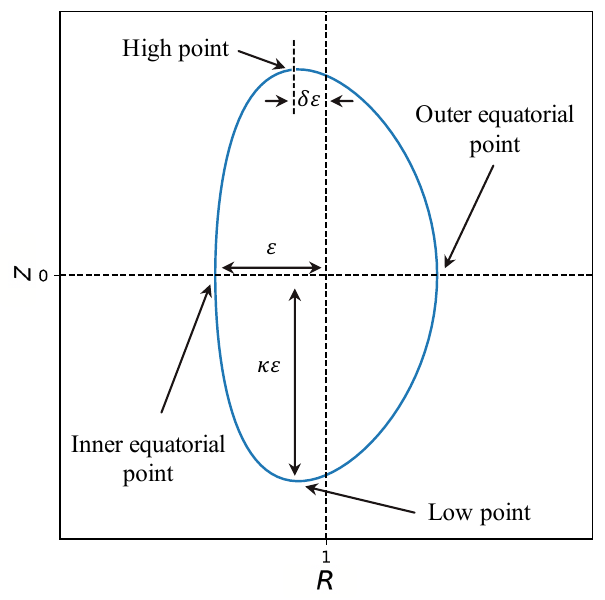}
    \caption{Elongated-D geometry from Cerfon and Freidberg.~\cite{cerfon2010one}}
    \label{fig:elongated_D}
\end{figure} 

\subsection{Solov'ev Equilibria}
In order to evaluate the performance of PINNs, this work explores a range of analytic GSE solutions.~\cite{solov1968theory} Although the general solution for the non-linear partial differential equation in \eqref{eq:full_gse} must be computed numerically, certain choices of $p(\psi)$ and $F(\psi)$ allow for an analytical solution: Solov'ev profiles, in which $p(\psi)$ and $F^2(\psi)$ are both linear functions of $\psi$:
\begin{align}
\label{eq:solovev_gse_unscaled}
    \mu_0 r^2\frac{dp(\psi)}{d\psi} &= -Cr^2 \notag\\
    \frac{1}{2}\frac{dF^2(\psi)}{d\psi} &= -A
\end{align}
where $A$ and $C$ are constants. After rescaling with the major radius of the plasma, $R_0$ and an arbitrary constant, $\psi_0$, via the definitions $R = r/R_0$, $Z = z/R_0$, $\psi = \psi_0 \Psi$, and $\psi_0 = R^2_0(A+CR^2_0)$,~\eqref{eq:full_gse} can be written
\begin{align}
\label{eq:sol_gse}
    \Psi_{RR} - \frac{1}{R}\Psi_R + \Psi_{ZZ} - PR^2 - (1-P) = 0,
\end{align}
where $P = CR_{0}^2/(A+CR_{0}^2)$ is the strength of the linear pressure profile.~\cite{cerfon2010one}.

\subsection{Fixed Boundary Condition}
We consider a toroidal, elongated ``D''-shaped geometry with a major radius $R_0$, as in Fig. ~\ref{fig:elongated_D}. The boundary can be defined by the following parametric equations:
\begin{align}
\label{eq:boundary_parametric_eq}
        R &= 1 + \varepsilon\cos(\tau + \arcsin(\delta)\sin(\tau)), \notag\\
        Z &= \varepsilon\kappa\sin(\tau),
\end{align}
where $\varepsilon = a/R_{0}$ is the inverse aspect ratio, $\kappa$ is the elongation, $\delta$ is the triangularity, and $\tau \in [0, 2\pi)$.

In the fixed-boundary scenario, the boundary condition is defined as
\begin{align}
\label{eq:fixed_bc}
    \Psi(R, Z)|_{\partial\mathcal{D}} = 0,
\end{align}
where $\partial\mathcal{D}$ is the boundary from~\eqref{eq:boundary_parametric_eq}.

\subsection{Analytic Solutions}
In order to find analytic solutions with Solov'ev profiles, we use a method from Cerfon et al.~\cite{cerfon2010one} With the prescribed boundary conditions and Solov'ev profiles, the Grad-Shafranov equation in~\eqref{eq:full_gse} can be solved analytically using a sum of the homogeneous and particular solutions, where
\begin{align}
\label{eq:analytic_solutions}
    &\Psi = \Psi_H + \Psi_P \\
    &\Psi_P = \frac{PR^4}{8} + (1-P)\frac{R^2}{2}\log(R) \notag\\
    &\Psi_H \approx \sum_{i=1}^7c_i\Psi_i. \notag
\end{align}
The $\Psi_i$ are up-down symmetric functions of $(R, Z)$, and can be found in Cerfon and Friedberg~\cite{cerfon2010one} or Zheng et al.~\cite{Zheng_2018} The coefficients $c_i$ are determined from seven point-wise boundary conditions specified in Cerfon et al. With the seven geometric constraints, we have seven requirements that uniquely determine $c_1, ..., c_7$. 
In principle, for the $\Psi=0$ contour to coincide with a specific boundary curve such as (\ref{eq:boundary_parametric_eq}) exactly, an infinite number of basis functions would need to be included, rather than seven. 
However, in practice this truncated analytic solution is very accurate for many choices of the boundary conditions. To be more precise, the relative error between the boundary shape prescribed by (\ref{eq:boundary_parametric_eq}) and the boundary shape of the Cerfon analytic solutions is on the order of $10^{-5}$ for the examples that will be considered here. Given $\nabla\Psi \sim 0.1$ in typical solutions, the difference of $\sim 10^{-5}$ in the boundary location translates to a difference of $\sim 10^{-5} * 0.1 \approx 10^{-6} = 10^{-4}\%$ in $\Psi$ locally. We conclude that the boundary of the Cerfon analytic solution is close enough to (\ref{eq:boundary_parametric_eq}) for (\ref{eq:analytic_solutions}) to be used as analytic solutions, and the difference between the the boundary shapes is subdominant to other solution errors that will be discussed in the paper.

\begin{figure}[t!]
    \centering
    \includegraphics[width=0.8\linewidth]{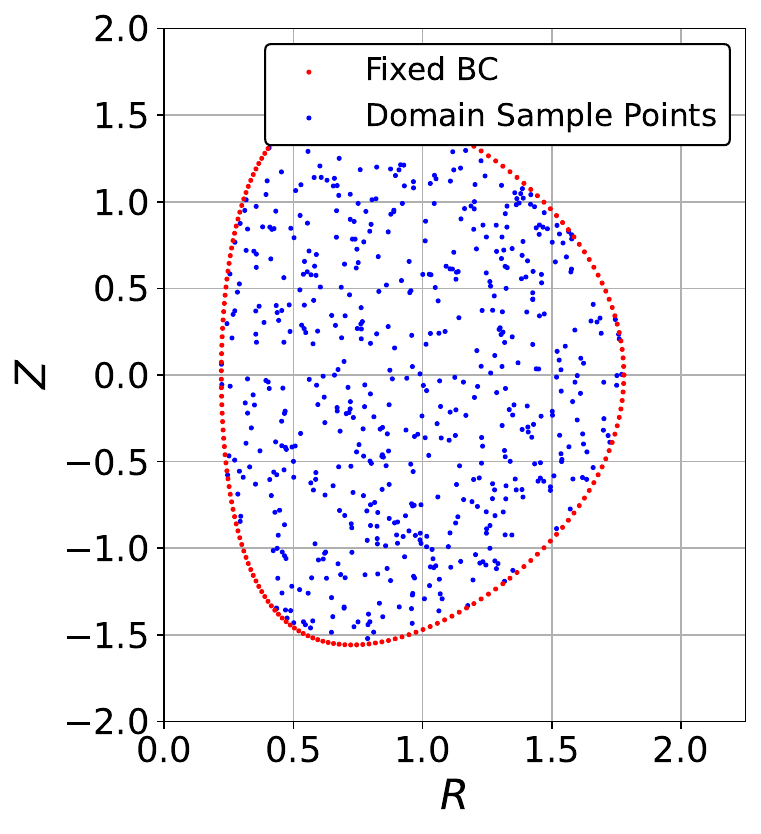}
    \caption{500 collocation points inside the domain where a governing equation should be satisfied and 100 collocation points on the boundary are plotted. }
    \label{fig:col_points}
\end{figure} 

\begin{figure*}[t!]  
    \centering
    \includegraphics[width=0.9\linewidth]{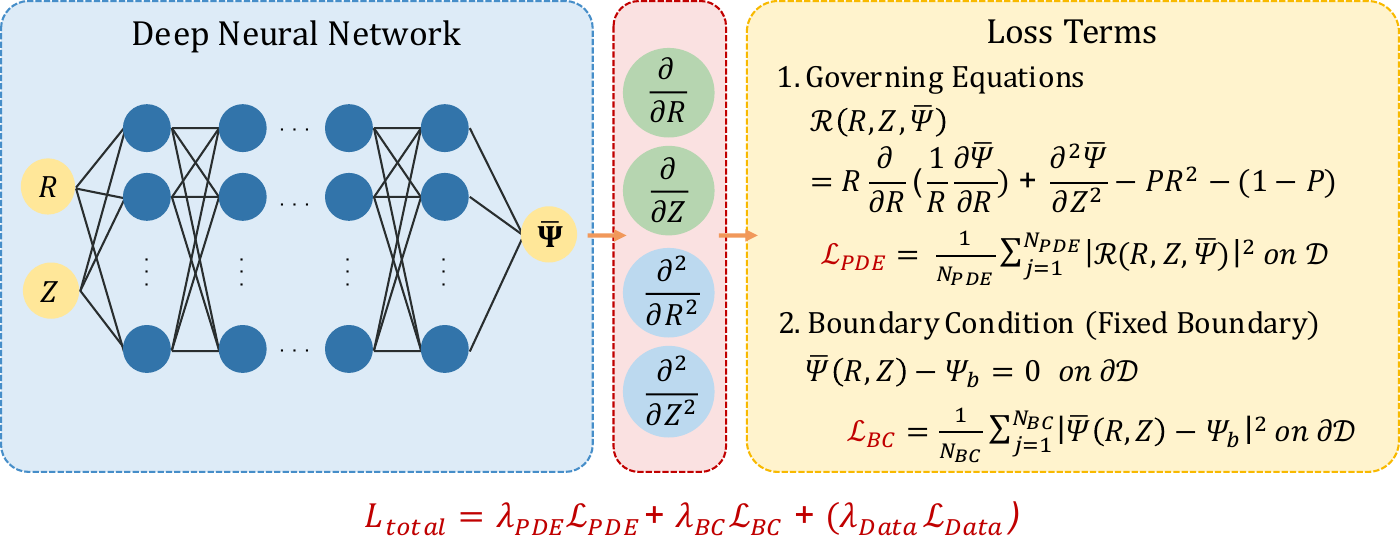}
    \caption{PINN Architecture for the Grad-Shafranov Equation}
    \label{fig:PINN_Arch}
\end{figure*}  

\section{Physics-Informed Neural Networks}
PINNs use a neural network as a universal function approximator~\cite{hornik1989multilayer} to calculate the solution of a PDE given independent variables of the PDE. Using automatic differentiation, residuals are calculated from the governing equation, boundary conditions,  and (possibly) data. 
The residuals are consolidated into a loss function with weights.~\cite{raissi2019physics} Fig. ~\ref{fig:PINN_Arch} describes the full PINN architecture used for this work.

Traditionally, feed-forward neural networks (FNNs) are used because of their simplicity and effectiveness; the neurons of adjacent layers are fully connected, and outputs of each layer are fed forward as the inputs to the next layer. FNNs map an input layer $\bm{z}_0 \in \mathbb{R}^{n_0}$ to the output $\bm{z}_{L} \in \mathbb{R}^{n_L}$, If $L > 1$ the network is considered ``deep.'' For the standard PINNs employed in this work, $\bm z_0 = (R, Z)$, $n_0 = 2$, $n_L = 1$ and $z_L$ is the neural network approximation of the poloidal flux function, $\bar{\Psi}$. The layers between the input and output layers are denoted hidden layers $\bm{z}_i$, $i = 1,...,L-1$, and
\begin{align}
    \bm{z}_i = \sigma_i(\bm{W}_i\bm{z}_{i-1} + \bm{b}_i).
\end{align}
where $\bm{W}_i \in \mathbb{R}^{n_{i}\times n_{i-1}}$ and $\bm{b}_i \in \mathbb{R}^{n_i}$ are the weight matrix and the bias vector, the subscript $i$ denotes the index of the layer, and $\sigma_i(·)$ is an activation function acting element-wise. The activation function can be any of the common varieties, e.g., sigmoids, rectified linear units (ReLU), and tanh functions. Different activation functions are explored in this work, and we found that ReLU does not work for solving the Grad-Shafranov equation since the solution needs to be twice differentiable. During training, both PDE and boundary condition residuals in a loss function are calculated by applying automatic differentiation to our differentiable surrogate model $\overline{\Psi}$. 
After the training, the weights, bias, and activation function at each layer are determined, and the output prediction $\overline{\Psi} = z_L$ can be efficiently calculated from any given input vector $\bm{z}_0$ (i.e., coordinates).

Since we are concerned with the solution to the GSE, we need to specify the equation and an appropriate set of boundary conditions on the boundary $\partial \mathcal{D}$: 
\begin{align}
    \Delta^* \Psi(R, Z) &= 
    \mathcal{N}(\Psi, R, Z), \\
    B(\Psi,R_{b},Z_{b}) &= 0, \quad (R_{b},Z_{b})\in \partial \mathcal{D}
\end{align}
where $\Delta^*$ and $\mathcal{N}$ is defined as the following:
\begin{align}
    \Delta^* \Psi(R, Z) &= \Psi_{RR} - \frac{1}{R}\Psi_R + \Psi_{ZZ} \\
    \mathcal{N}(\Psi, R, Z) &= PR^2 + (1-P).
\end{align}
$\mathcal{N}$ is for now an arbitrary function of $\Psi$ and the coordinates, determined by the pressure and poloidal current profiles, and $B(\Psi, \bm{x})$ denotes the boundary condition for $\Psi$ not a magnetic field. Then the PINN approach with ``soft'' boundary condition can be represented as a minimization problem with a boundary condition as a penalty term:

\begin{align}
    \label{eq:soft_constraints_min}
    \argmin_{\bm{W}, \bm{b}} \|\Delta^*\overline{\Psi} - \mathcal{N}(\overline{\Psi}, R, Z)\|^2_2 + \lambda_1\|B(\overline{\Psi},R_{b},Z_{b})\|^2_2,
\end{align}
where $\lambda_1$ is a hyperparameter weighing the tradeoff between matching the PDE in the volume and matching the boundary conditions.

For some problems, imposing ``hard'' instead of ``soft'' boundary condition constraints can improve performance.~\cite{sun2020surrogate,lu2021physics} Therefore, both soft constraints and hard constraints for boundary conditions are explored. The hard constraint can be effectively imposed by using a function $G(R,Z)$ that adheres strictly to the boundary condition. The function $G(R,Z)$ can be any smooth function that goes to zero at the boundary. It can be found with two different methods. We can find an analytical form from a given geometry or train a separate neural network that vanishes at the boundary, and multiply the two network outputs. Either way, the final neural network with hard constraints,$\overline{\Psi}(R,Z)_{hc}$,  becomes:
\begin{align}
    \label{eq:hard_constraints_NN_form}
    \overline{\Psi}(R,Z)_{hc} = G(R,Z)\overline{\Psi}(R,Z). 
\end{align}

The drawbacks of employing a hard constraint approach include potentially lower accuracy for certain problems and the limitation that it can only be implemented when there is a functional representation of the boundary.

Note that, since the PDE is known, we can generate MHD equilibria without any actual data from a simulation or experiment. However, it often improves the results to include a set of data points $(R_{data}, Z_{data})$, either numerical or experimental data, where we would like the measured and computed solutions to match,
\begin{align}
\label{eq:pinn_full}
    \argmin_{\bm{W}, \bm{b}} \|\Delta^*\overline{\Psi} - \mathcal{N}(\overline{\Psi}, R,Z)\|^2_2 + \lambda_1\|B(\overline{\Psi},R_{b},Z_{b})\|^2_2 \\
    + \lambda_2\|\tilde{\Psi}(R_{data}, Z_{data}) - \overline{\Psi}(R_{data},Z_{data})\|_2^2.  \notag
\end{align}
$\tilde{\Psi}(R_{data},Z_{data})$ denotes measurements of the poloidal flux function from a simulation or experiment at given points $R_{data}$ and $Z_{data}$; in the case of an experiment, $\tilde{\Psi}$ should be understood to be an approximation of the GSE corrupted by measurement noise and any non-equilibrium dynamics. Although we solve a number of fixed-boundary problems in Sec~\ref{sec:forward_problem}, we omit any data points to illustrate that these methods can reproduce approximate MHD equilibria from the minimization of the PDE residual and boundary conditions alone. Unlike data points that are from either a simulation or experiment, collocation points are points in input space where the residuals are calculated. Collocation points are different from data points in that there is no need for experiment or simulation to collect and label the data. An example of collocation points used in this work is shown in Fig.~\ref{fig:col_points}.

\section{Implementation and Training}
We will first use a standard FNN with three hidden layers and 20 nodes in each hidden layer. The inputs are the coordinates $(R, Z)$ and the output is $\overline{\Psi}$. To try to avoid local optima, the Adam optimizer~\cite{kingma2014adam} with learning rate $\alpha_l = 10^{-3}$ was used for the first 100 iterations and then L-BFGS-B~\cite{zhu1997algorithm} is used. We used 100 and 500 boundary and domain collocation points, respectively. In Appendix B, the parameter space is explored by varying the size of the model, activation functions, and optimizers, in order to map various trade-offs such as between reconstruction error and computational speed.We demonstrate that approximating the solution to the two-dimensional Grad-Shafranov equation with linear profiles can be done with absolute errors of no more than one percent. We observe no significant effect on our results when using larger networks and adjusting the hyperparameters. In the subsections that follow, we will explain how we use PINN in various scenarios.

\subsection{Simple Equilibria}
The first equilbria considered have smooth cross sections described by \eqref{eq:boundary_parametric_eq}. The domain collocation points are randomly generated using Latin hypercube sampling. These points are where the residual of the GS PDE~\eqref{eq:sol_gse} is calculated. To be clear, these collocation points and residuals are not obtained via running other equilibrium solvers. The collocation points are merely points where the residuals will be calculated. Equation \eqref{eq:boundary_parametric_eq} is used to apply a Dirichlet boundary condition (i.e. $\Psi_{b} = 0 $) along the cross-section boundary. The residuals of the boundary loss will be calculated according to ~\eqref{eq:gen_bound_loss_term}. The calculated residuals are then used to calculate the loss function. 
\begin{align}    
    \label{eq:gen_tot_loss_term}
    L_{total} &= L_{pde} + \lambda_{1}L_{bc},
     \\
    \label{eq:gen_pde_loss_term}
    L_{pde} &= \frac{1}{N_{pde}}\sum_{j=1}^{N_{pde}} \| \Psi_{RR} - \frac{1}{R}\Psi_R + \Psi_{ZZ} - PR^2 - (1-P) \|^2_2,
     \\
    \label{eq:gen_bound_loss_term}
    L_{bc} &=  \frac{1}{N_{bc}}\sum_{j=1}^{N_{bc}} \|\Psi(R_j,Z_j)\|^2_2.
\end{align}
We found that the highest accuracy is obtained with $\lambda_{1} = 100$. 

\subsection{Equilibria with a Divertor/X-Point}
Equilibria with an X-point can be calculated similar to the simple equilibria examples. Since we do not have a convenient parametric equation for the boundary like~\eqref{eq:boundary_parametric_eq}, the boundary is obtained through an analytical solution. We can then generate collocation points both in the domain and at the boundary.

\subsection{Hard Constraints}
In the method shown in Figure ~\ref{fig:PINN_Arch}, we apply soft constraints of the boundary conditions through the loss $L_{bc}$ shown in ~\eqref{eq:gen_bound_loss_term}. Soft constraints on the boundary condition were used for simple- and divertor-equilibria cases. The hard constraints on the boundary conditions can be advantageous for various reasons. For some problems, it is shown that the hard constraints produce faster and more accurate solutions to a given PDE. It might be important in various diagnostic scenarios or in shape optimization to match the boundary condition exactly. 

For this work, we derived the following hard constraint function $G(R,Z)$  from ~\eqref{eq:boundary_parametric_eq}:
\begin{align}
    \label{eq:top_down_symmetry_Gfunction}
    G(R,Z) = |Z|-\varepsilon \kappa\sin{\left(\arccos{\left(\frac{|R|-1}{\varepsilon}\right)}-\arcsin(\delta)\frac{|Z|}{\varepsilon\kappa}\right)}.
\end{align}

\subsection{Parametric-PINN}
\begin{figure}
    \centering
    \includegraphics[width=\linewidth]{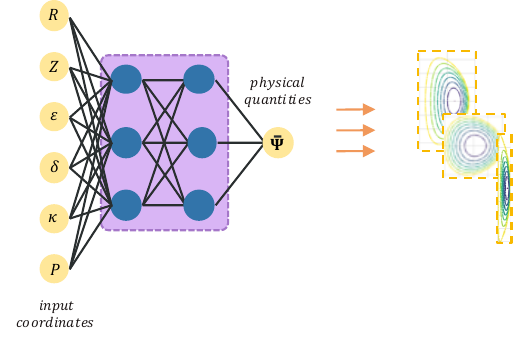}
    \caption{Modified input space for parametric PINN. One trained parametric-PINN can generate equilibria for different system configurations.}
    \label{fig:Parametric_PINN_Arch}
\end{figure}

So far, each trained PINN model had fixed shape parameters, so the PINN can only produce solutions for one specific boundary configuration. However, it can be helpful to be able to interpolate and extrapolate for various boundary configurations with one pre-trained model. This can allow us to solve inverse problems, such as shape optimization for reactor design. When expanding to the parametric-PINN framework, the input space includes additional variables: strength of the pressure profile $P$, inverse aspect ratio $\varepsilon$, elongation $\kappa$, and triangularity $\delta$. This extended input space requires modifications to the training and implementation procedures, as depicted in Fig.~\ref{fig:Parametric_PINN_Arch}. Instead of only the spatial ($R$,$Z$) coordinates, the input to the PINN is now a 6-dimensional vector [$R$, $Z$, $P$, $\varepsilon$, $\kappa$, $\delta$] and the neural network output is $\overline{\Psi}(R,Z,P,\varepsilon, \kappa,\delta)$. In principle, one could include other parameters that affect the equilibria as additional variables in the input layer. 

For the PDE residuals, we use the Grad-Shafranov equation with varying $P$. Soft boundary conditions are used, in the same way as in the non-parametric case. The implementation of the trained parametric-PINN model can be realized by feeding it with the required 6-dimensional inputs, thereby facilitating accurate and efficient solutions to the Grad-Shafranov equation under different plasma scenarios characterized by varied pressure profiles, aspect ratios, elongations, and triangularities.

\section{Results}\label{sec:forward_problem}
\begin{table*}
    \centering
    \setlength{\extrarowheight}{2pt}
    \begin{tabular} {
    >{\hspace{5pt}}l<{\hspace{5pt}}
    >{\hspace{5pt}}l<{\hspace{5pt}}
    >{\hspace{7pt}}l<{\hspace{7pt}}
    >{\hspace{7pt}}l<{\hspace{7pt}}
    >{\hspace{7pt}}l<{\hspace{7pt}}
    >{\hspace{7pt}}l<{\hspace{7pt}}
    }
    \toprule
    \multicolumn{2}{c}{Configuration}                   \\
    \cmidrule(r){1-2}
    Devices     & $P$     & average error(\%) & max error(\%) & training time & inference time \\
    \midrule
    ITER      &  0  &  0.015  &  0.162  &  20s  &  10ms   \\
    NSTX      &  0  &  0.032  &  0.142  &  22s  &  15ms   \\
    Spheromak &  0  &  0.027  &  0.204  &  36s  &  12ms   \\
    ITER      &  1  &  0.021  &  0.136  &  17s  &  11ms \\
    NSTX      &  1  &  0.026  &  0.186  &  32s  &  13ms \\
    Spheromak &  1  &  0.075  &  0.325  &  37s  &  10ms \\
    FRC       &  1  &  0.057  &  0.214  &  27s  &  17ms \\
    \bottomrule
    \end{tabular}
    \caption{\raggedright Performance of the PINNs for the baseline configurations explored in this work.}.
    \label{tabele:configurations_results}
\end{table*}

We first demonstrate the accuracy and effectiveness of PINNs for solving the GSE under two different Solov'ev profiles for four different devices: ITER, NSTX, a spheromak, and an FRC (field-reversed configuration). In addition to these configurations, we also investigated variations with divertors, further expanding the applicability of our method. We benchmark the PINN-based solutions against the analytical solutions in~\eqref{eq:analytic_solutions}, which serve as a reliable reference for our study. 

In addition, we examine the impact of using hard constraints instead of soft constraints in the PINN model. Our findings show that hard constraints decrease computational complexity and relative error, as they satisfy an analytical representation for the boundary by definition, eliminating the need to evaluate collocation points for each boundary.

Lastly, we present preliminary results exploring parametric PINN, where a single neural network can be trained for various devices with different geometries. Our model suffered from the curse of dimensionality, as the model's larger input space was associated with an exponentially increased number of boundary collocation points. Nevertheless, we were able to train a reasonably effective surrogate for MHD equilibria parametrized by the shape parameters, with promise for inverse problems. 

\subsection{Baseline Equilibria with Different Configurations}
\begin{figure}
    \centering
    \includegraphics[width=\linewidth]{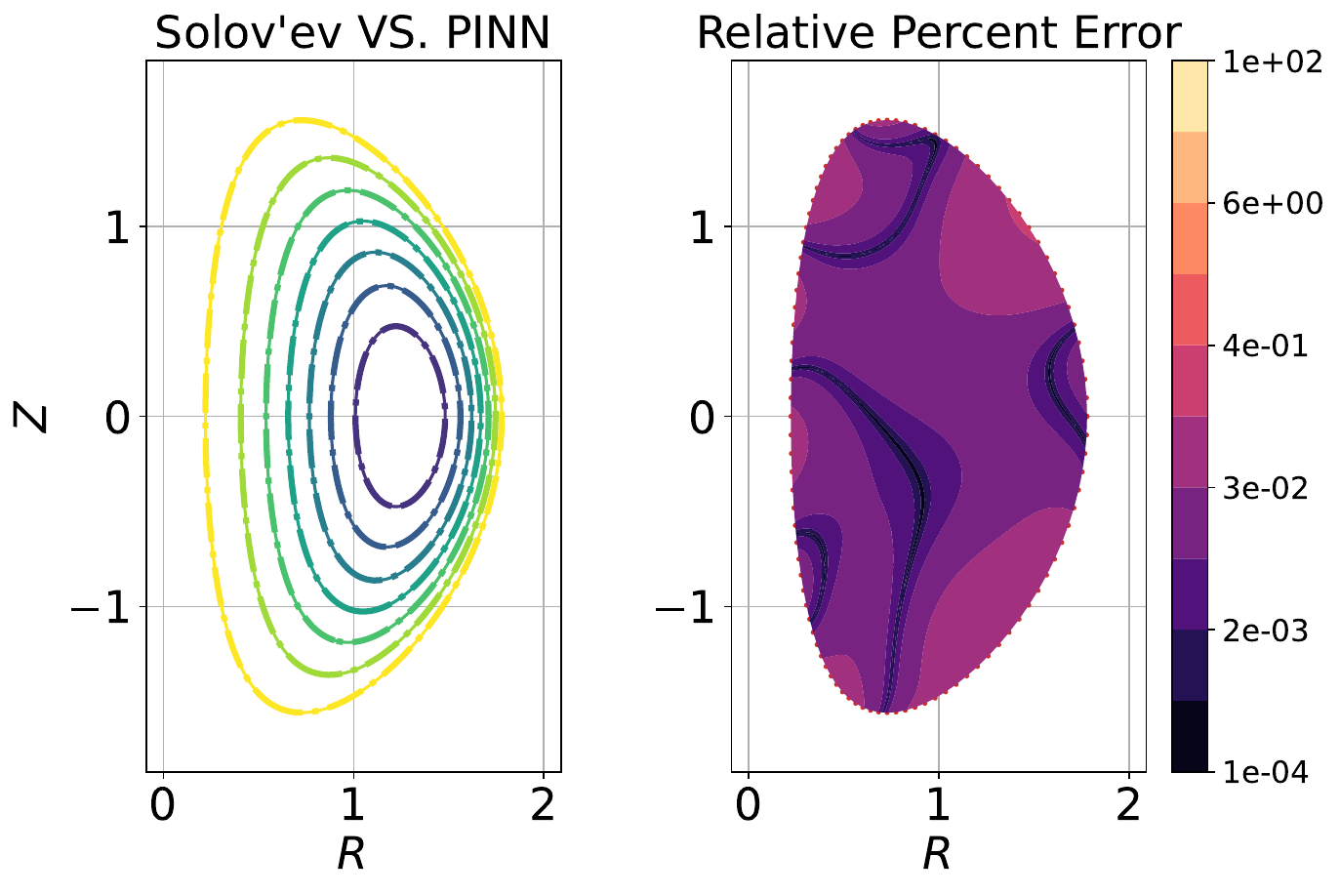}
    \caption{MHD equilibrium calculated using NSTX geometry with finite beta. The plot on the left shows the solid and dot-dashed curves that indicate the PINN and analytic solutions respectively.}
    \label{fig:reconstruction_result_baseline_nstx}
\end{figure}

The seven different configurations considered in this study are displayed in Table ~\ref{table:configurations} in Appendix A. By assessing the performance of PINNs in these diverse scenarios, our objective is to demonstrate the robustness and versatility of our approach to solving GSE in various magnetic confinement devices. As one example, NSTX geometry with finite beta is shown in Fig.~\ref{fig:reconstruction_result_baseline_nstx}. 

We evaluate the performance of PINNs by analyzing several important figures of merit, such as relative error (both average and maximum) and computational time (both training and inference time). These metrics provide insight into the accuracy, efficiency, and overall effectiveness of our approach to modeling the different plasma configurations. Our main metric for performance is the relative error as a percent:
\begin{align}
    \label{eq:relative_error}
    E(R,Z) = 100\% \frac{|\overline{\Psi}-\Psi^{*}|}{|\overline{\Psi_a}|},
\end{align}
where 
$\overline{\Psi}$ is a neural network surrogate for the flux function, $\Psi^{*}$ is the analytic solution, and $\overline{\Psi}_a$ is the maximum value of the neural network surrogate for the flux function, i.e. the value at the magnetic axis. Table ~\ref{tabele:configurations_results} shows the ability of our method to accurately and effectively solve the GSE for a wide range of plasma geometries and magnetic field conditions. The spatially averaged relative errors were around $~0.02\%$, which is low enough to closely reproduce the correct equilibrium. Notably, increasing the number of layers or nodes in the neural network  only marginally reduces these errors further, unlike traditional equilibrium solvers, which can find equilibria to high numerical precision. This loss of accuracy compared to traditional numerical solvers is inevitable,~\cite{markidis2021old} so PINNs should be used for problems where computational speed or inverse solves are more important than obtaining equilibria to high numerical precision. 

\begin{figure}
    \centering
    \includegraphics[width=\linewidth]{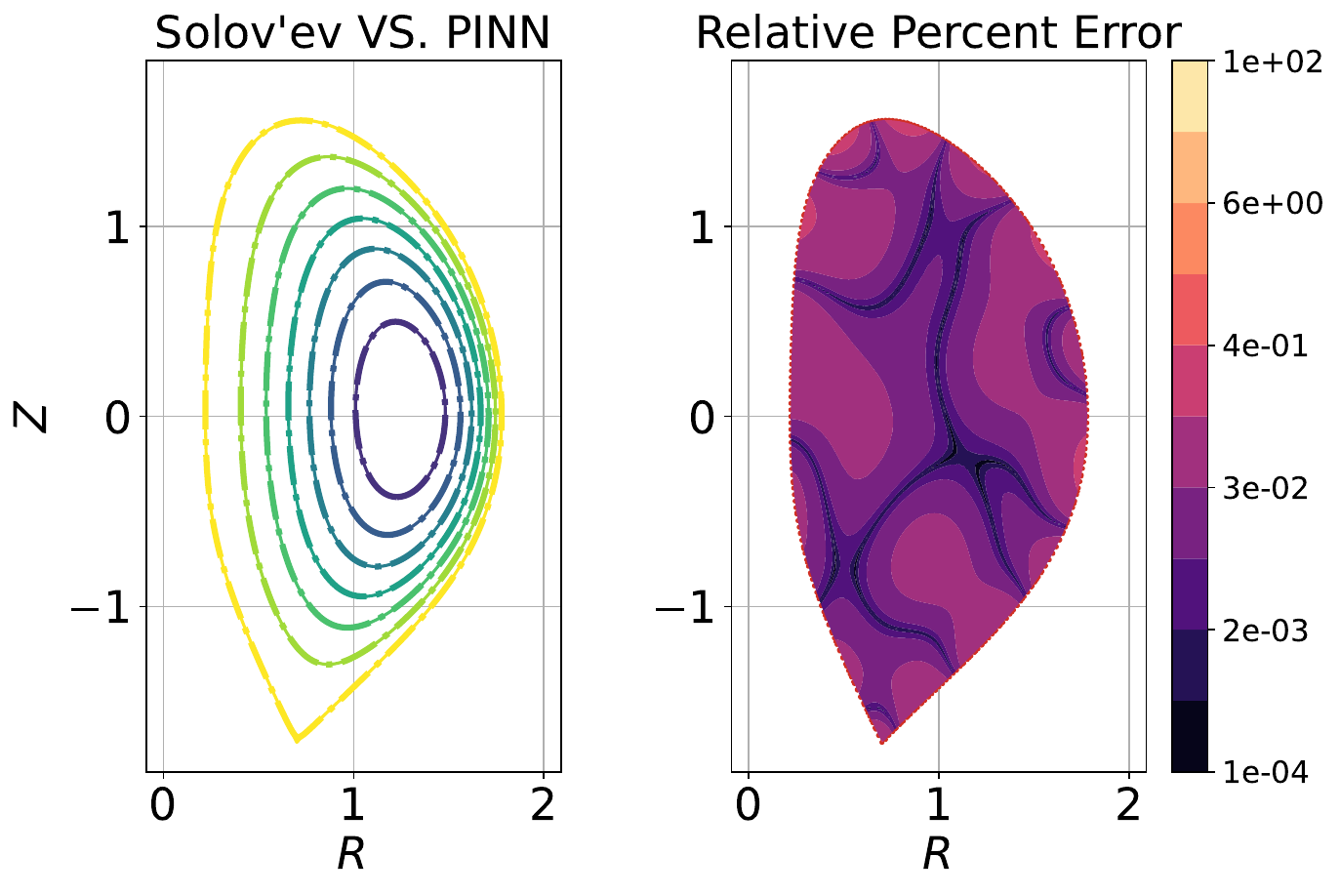}
    \caption{MHD equilibrium calculated using NSTX geometry with finite beta with a lower divertor.}
    \label{fig:NSTX_singlenull}
\end{figure}

\subsection{Configurations with a divertor}
For a diverted configuration with X-point(s) in the separatrix, the same neural network architecture is used, with some modification in the boundary conditions. Similar to the simple equilibria case, we implemented these configurations with a fixed boundary condition. Since we do not an explicit analytic expression for the boundary shape, we numerically located points on the boundary of the analytic solution for $\Psi$, and used the resulting points for the boundary loss term. As shown in Fig.~\ref{fig:NSTX_singlenull}, the accuracy and training time were comparable to the baseline equilibria configurations without an X-point.
\subsection{Hard vs. Soft Constraints}

\begin{figure}
    \centering
    \includegraphics[width=\linewidth]{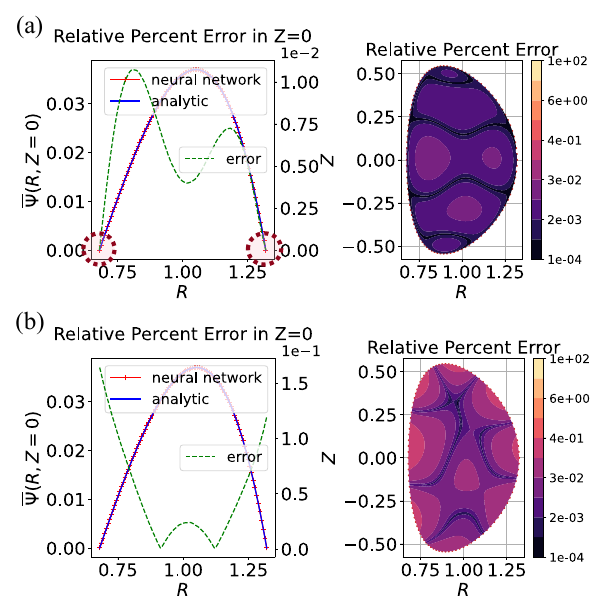}
    \caption{Comparing soft and hard constraints for ITER configuration. The model shown in (a) is a hard constraint model and in (b) is a soft constraint model. For the hard constraint case (a), the error is zero on the boundary, verifying the hard constraint.}
    \label{fig:soft_vs_hard}
\end{figure}

PINNs for configurations without an X-point were trained using both hard and soft constraints, and the figures of merit mentioned earlier were compared: equilibrium reconstruction error (both average and maximum) and computational time (both training and inference time).
The results with hard constraints show 2 to 10 times more accurate equilibria reconstructions with faster training time. Qualitatively, we can see that the error at the boundary is zero by~\eqref{eq:hard_constraints_NN_form} and can be visually confirmed in Fig.~\ref{fig:soft_vs_hard}. As mentioned before, even though the Cerfon analytic solution does not align exactly with~\eqref{eq:boundary_parametric_eq}, using the Cerfon analytic solution as a comparison is justified as the relative error between the two is subdominant, i.e. $\sim 10^{-4}\%$.

\subsection{Parametric-PINNs}
We present a preliminary work on parametric-PINNs in which a single neural network is trained for various devices with different geometries. A single trained model can interpolate and even extrapolate for various boundary configurations. For training, we varied each parameter (e.g. $P$, $\varepsilon$, $\kappa$, and $\delta$) as follows: $P = [0.00, 1.00]$, $\varepsilon = [0.12, 0.52]$, $\kappa = [1.25,2.75]$, and $\delta = [-0.5, 0.5]$. For each parameter, we sampled nine points in the given range. The model had four layers with 30 nodes in each layer. 

\begin{figure*}
    \centering
    \includegraphics[width=\linewidth]{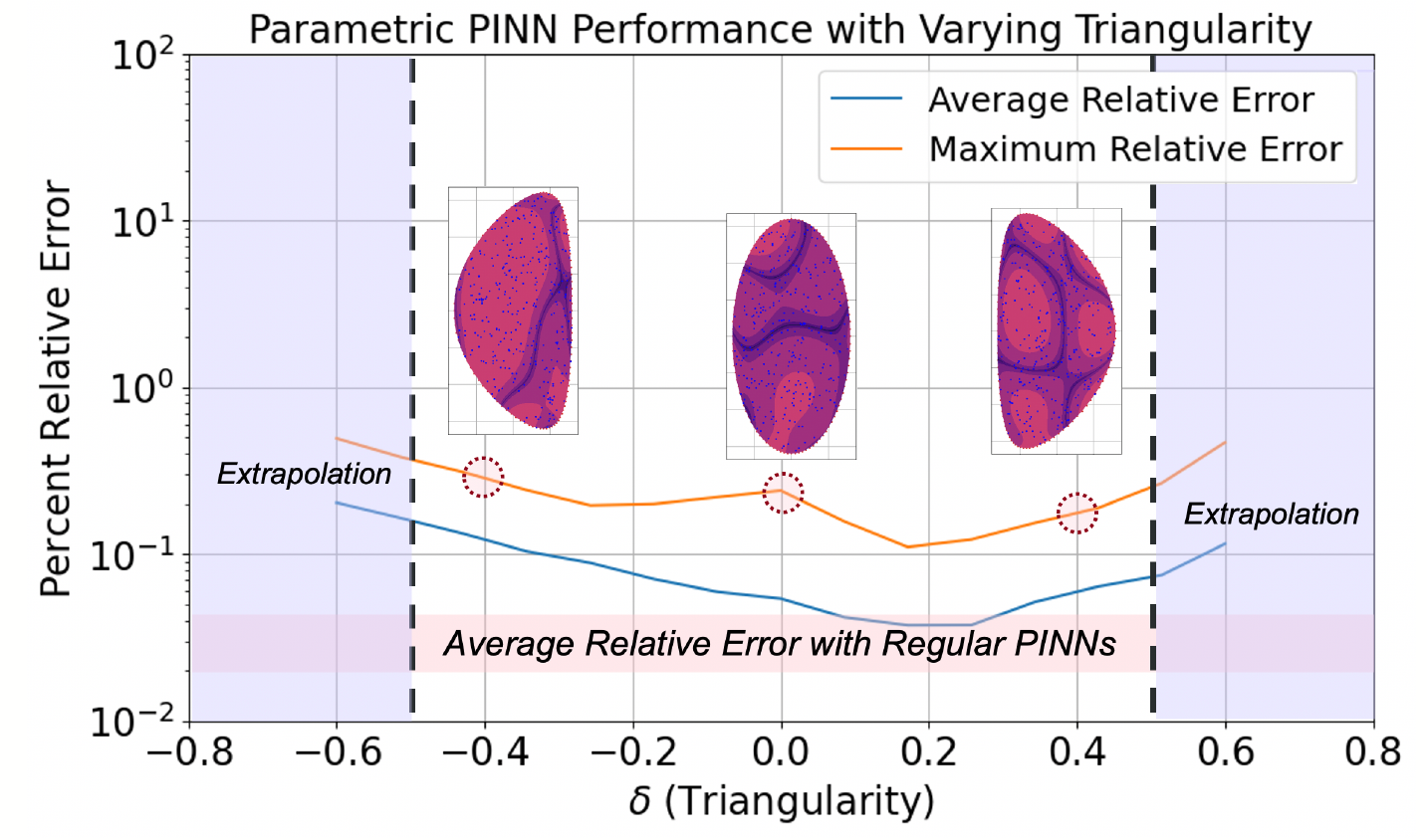}
    \caption{Relative percent error of a pretrained parametric-PINN for various triangularities. At the worst interpolation regions, it is an order of magnitude worse than a regular PINN. The colorbars of the configurations are the same as the colorbars in Figures 5-7.}
    \label{fig:parampinn_results_del}
\end{figure*}

The training for the parametric-PINNs took longer than the regular PINNs (i.e. four hours vs. 20 seconds) and the reconstruction error increased by up to an order of magnitude as shown in Fig. \ref{fig:parampinn_results_del}. The inference times are comparable to regular PINNs at around 10ms. With 6D input instead of 2D input space for the model, the number of collocation points exponentially increased. The number of collocation points scales as $N^{D}$ where $D$ is the number of dimensions and $N$ is the number of points sampled in each dimension (assuming for simplicity one does uniform sampling). For example, we have four dimensions (i.e. $P$, $\varepsilon$, $\kappa$, $\delta$) with nine points sampled in each dimension (i.e. for triangularity: -0.5, -0.375, -0.25, -0.125, 0.0, 0.125, 0.25, 0.375 and 0.5). So, if we have a 100 boundary collocation points (R, Z) for each configuration (i.e. at a fixed $P$, $\varepsilon$, $\kappa$, $\delta$), the total number of boundary collocation points for parametric-PINNs will be $100 \times 9^{4} = 656,100$. For the domain collocation points, we also used $656,100$ pseudo-randomly generated points.

A preliminary result on the trained parametric-PINNs is shown in Fig. ~\ref{fig:parampinn_results_del} which shows the variation in $\delta$ from a single trained model instead of three different models. The three other dimensions (i.e. $P$, $\varepsilon$, and $\delta$) can also be varied in the same model without re-training, but results are not show here as the trend is similar. The error gradually gets worse as we extrapolate further from the trained region. 

Parametric PINNs have the potential to be as accurate as regular PINNs with further hyperparamater tuning and other techniques such as hard constraints and adaptive sampling methods.~\cite{tang2023pinns, wu2023comprehensive} 
Exploring hard constraints for boundary conditions in parametric PINNs is beyond the scope of this work, but hard constraints could be advantageous by providing a workaround for the exponentially increasing number of collocation points. Furthermore, given the high-dimensional input space, strategies for efficient sampling of collocation points (both interior and boundary points) are also imperative in future work.

\section{Summary and Conclusions}
In this study, we have illustrated a comparison between the analytical solution and PINN in various fusion devices, such as ITER, NSTX, a spheromak, and an FRC. Although the PINNs approach may not be the optimal choice for achieving the highest precision compared to modern numerical MHD equilibrium codes, it offers several advantages. Specifically, the PINN approach (1) eliminates the need for complicated PDE discretization and solution schemes, requiring only the coding of the PDE in symbolic language, and (2) provides flexibility through automatic differentiation, enabling its use on various geometries and plasma devices without mesh generation.

The potential of PINNs to incorporate physical constraints into the loss function, as well as its flexibility, implies that it could be a valuable tool for MHD equilibrium reconstruction and optimization in a variety of devices with further improvements. PINNs are also known to excel in addressing inverse problems, a topic that should be explored in future work. 
\begin{table}
  \centering
  \setlength{\extrarowheight}{2pt}
  \begin{tabular} {
    >{\hspace{5pt}}l<{\hspace{5pt}}
    >{\hspace{5pt}}l<{\hspace{5pt}}
    >{\hspace{7pt}}l<{\hspace{7pt}}
    >{\hspace{7pt}}l<{\hspace{7pt}}
    >{\hspace{7pt}}l<{\hspace{7pt}}
    >{\hspace{7pt}}l<{\hspace{7pt}}
  }
    \toprule
    \multicolumn{2}{c}{Configuration}                   \\
    \cmidrule(r){1-2}
    Devices     & $P$     & $\epsilon$ & $\kappa$ & $\delta$ \\
    \midrule
    ITER      &  0.0    &  0.32  &  1.7  &  0.33   \\
    NSTX      &  0.0    &  0.78  &  2.0  &  0.35  \\
    Spheromak &  0.0    &  0.95  &  1.0  &  0.2  \\
    ITER      &  1.0  &  0.32  &  1.7  &  0.33  \\
    NSTX      &  1.0  &  0.78  &  2.0  &  0.35  \\
    Spheromak &  1.0  &  0.95  &  1.0  &  0.2   \\
    FRC       &  1.0  &  0.99  &  10.0 &  0.7   \\
    \bottomrule
  \end{tabular}
    \caption{\raggedright Shape parameters and pressure for all the configurations explored in this work. Here, 
    $P$ is the pressure parameter  in~\eqref{eq:sol_gse},
    $\epsilon = a/R_{0}$ is the inverse aspect ratio, $\kappa$ is the elongation, and $\sin(\alpha) = \delta$ is the triangularity.}
  \label{table:configurations}
\end{table}

Our work makes initial progress in employing PINNs for plasma physics applications, but we acknowledge that there is still room for improvement and further exploration. Future research should concentrate on refining the model by improving sampling methods for collocation points and broadening its applicability to a wider range of magnetic confinement devices. Although our preliminary parametric PINN approach encounters obstacles such as the curse of dimensionality, future work can investigate potential solutions, including separable PINNs,~\cite{cho2022separable} hard constraints, and adaptive sampling techniques.~\cite{tang2023pinns, wu2023comprehensive} This work can also expand to introduce general pressure and current profiles instead of the linear profiles shown in this work. Then, this work can expand to free-boundary equilibria, shape optimization for tokamaks, and 3D equilibria for stellarators, which have more complicated features such as magnetic islands.
\\

\section*{Acknowledgements}
We would like to acknowledge valuable discussions with Chris Hansen, Andrea Merlo, Timo Thun, and Thomas Sunn Pederson. 
This work was supported by the U.S. Department of Energy under contract No. DE-FG02-93ER54197. This research used resources of the National Energy Research Scientific Computing Center (NERSC), a U.S. Department of Energy Office of Science User Facility located at Lawrence Berkeley National Laboratory, operated under Contract No. DE-AC02-05CH11231 using NERSC award FES-ERCAP-mp217-2023.

\section*{Data Availability Statement}

The data that support the findings of this study are available from the corresponding author upon reasonable request.

\appendix
\section{Summary of configurations}
The seven different configurations considered in this study are displayed in Table~\ref{table:configurations} including their shape parameters and Solov'ev profile.

\section{Hyperparameter Analysis}
The parameter space of PINN architecture is explored by varying the activation functions, learning rate, number of iterations of Adam optimizer before using L-BFGS-B, and the size of the model, in order to map various tradeoffs such as between reconstruction error and computational speed. 
For each set of hyperparameters, we repeated the training 10 times with different random seeds for initialization. 

\begin{figure}
    \centering
    \includegraphics[width=\linewidth]{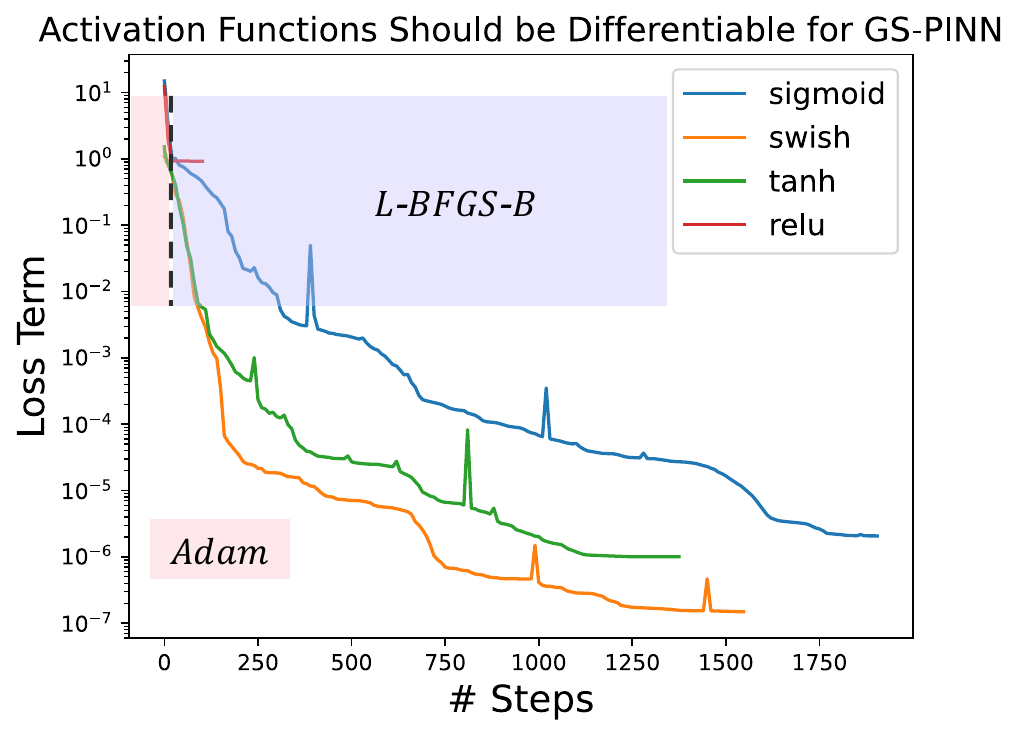}
    \caption{For these runs, 10 Adam steps were used and the swish function shows the best performance. ReLU fails to converge for all variations.}
    \label{fig:activation_hpt}
\end{figure}

For activation functions, there are four that were explored: sigmoid, swish, tanh, and relu:
\begin{align} 
    &\sigma(x) = \frac{1}{1+e^{-x}}, \\[10pt]
    &\text{swish}(x) = x \sigma(x),\\[10pt]
    &\text{tanh}(x) = \frac{e^{x} - e^{-x}}{e^{x} + e^{-x}},\\[10pt]
    &\text{ReLU}(x) = \max(0, x).
    \label{eq:activation_functions}
\end{align} 
The  sigmoid function is used as an activation function in binary classification problems, yielding a value between 0 and 1, which can be interpreted as a probability. However, sigmoid can suffer from the vanishing gradients problem, which can make training of deep networks challenging.~\cite{lecun2002efficient, wang2021understanding}

\begin{figure}
    \centering
    \includegraphics[width=\linewidth]{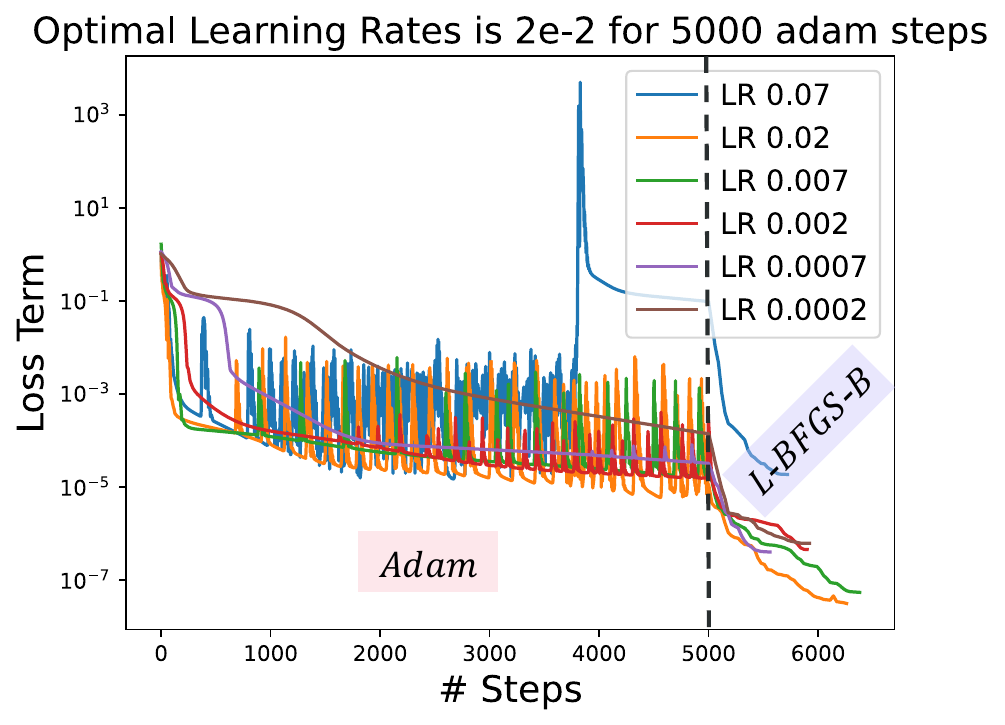}
    \caption{Learning rate of Adam optimizer of anything higher than 2e-2 starts to diverge due to taking big steps.}
    \label{fig:adam_lr_hpt}
\end{figure}

The hyperbolic tangent (tanh) function returns values that range from -1 to 1. The tanh function is similar to the sigmoid function but the mean of its output is closer to zero, which helps with converging the model during gradient descent.~\cite{lecun2002efficient}

The rectified linear unit (ReLU) returns the input directly if it is positive; otherwise, it evaluates to zero.~\cite{nair2010rectified, agarap2018deep} ReLU has become a default choice for many problems due to its simplicity and performance across multiple datasets and tasks. Its major benefit is that it solves the vanishing gradient problem, allowing models to learn faster and perform better. The issue with ReLU for this work is that the function is not differentiable at the origin. It is clearly shown in Fig. ~\ref{fig:activation_hpt} that models with ReLU do not converge.

\begin{figure}
    \centering
    \includegraphics[width=\linewidth]{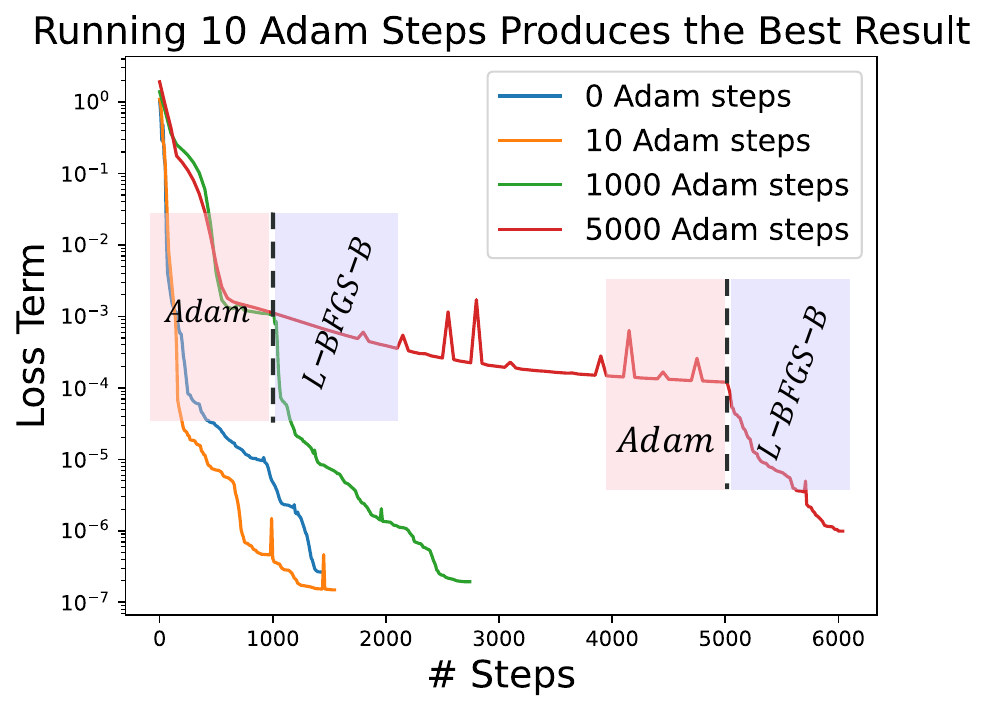}
    \caption{Running a smaller number of Adam optimizer before L-BFGS-B optimizer generally achieves higher accuracy and shorter computation time. However, fewer Adam steps cause greater uncertainties in the end result depending on the initialization.}
    \label{fig:adam_bfgs_hpt}
\end{figure}

Swish attempts to combine the best properties of ReLU and sigmoid functions. It is non-monotonic and this can help the model learn more complex patterns. It is smooth and differentiable, and unlike ReLU, it does not nullify negative input values.~\cite{ramachandran2017searching} 

\begin{figure}
    \centering
    \includegraphics[width=\linewidth]{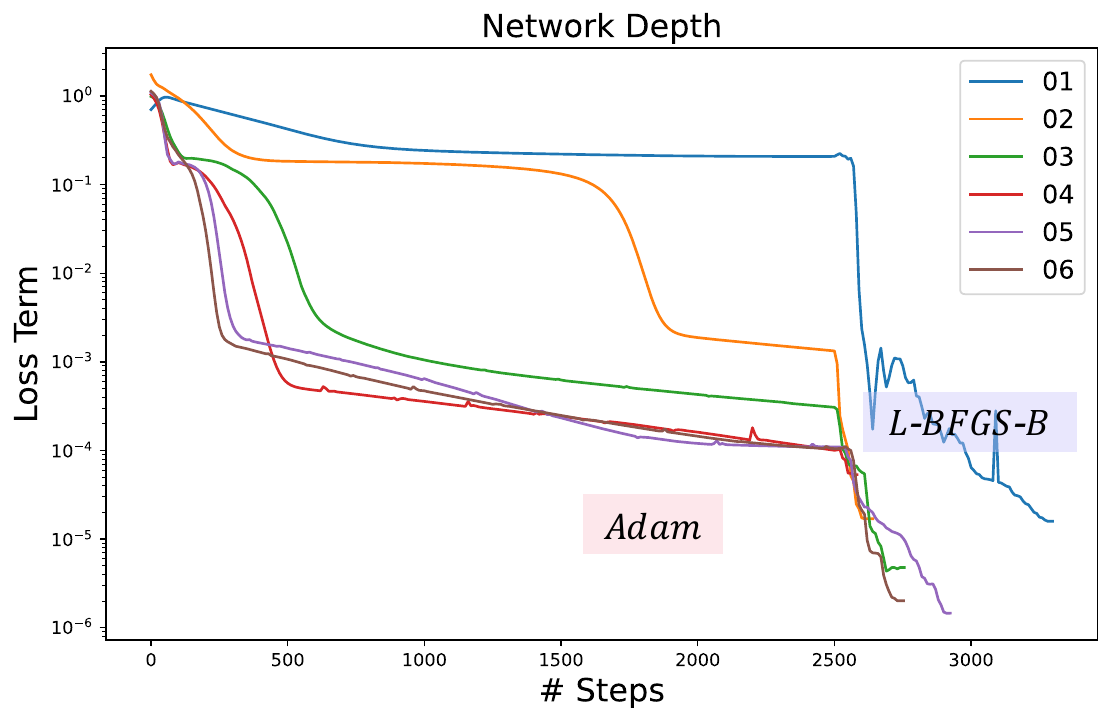}
    \caption{There is no significant improvement in accuracy for anything deeper than three layers.}
    \label{fig:depth_hpt}
\end{figure}

PINNs were trained with these four activation functions with a learning rate of 2e-2, with 100 Adam iterations followed by L-BFGS-B, network size of 3 hidden layers with 20 nodes each, 100 boundary points and 1000 domain points. As shown in Fig. ~\ref{fig:activation_hpt}, swish and tanh performs significantly better than the sigmoid function, and it is inconclusive whether tanh or swish is optimal. Models had a hard time converging with ReLU. This is probably because for the Grad-Shafranov equation, the model needs to be two times differentiable, where the ReLU function is not differentiable at the origin.

Next, the learning rate was explored by varying it from 7e-1 to 2e-4. The learning rate of the Adam optimizer is considered an important hyperparameter for effective neural network training.~\cite{wu2019demystifying} Depending on each model's loss landscape, an optimized learning rate can provide an effective training. If the learning rate is too high, the model may diverge whereas if the learning rate is too low it may take too long to converge.
As shown in Fig. ~\ref{fig:adam_lr_hpt}, for 1000 Adam steps the optimal learning rate is around 2e-2 and we see that the model starts to diverge at 7e-2. Note that the choice of these learning rates are arbitrary.

\begin{figure}
    \centering
    \includegraphics[width=\linewidth]{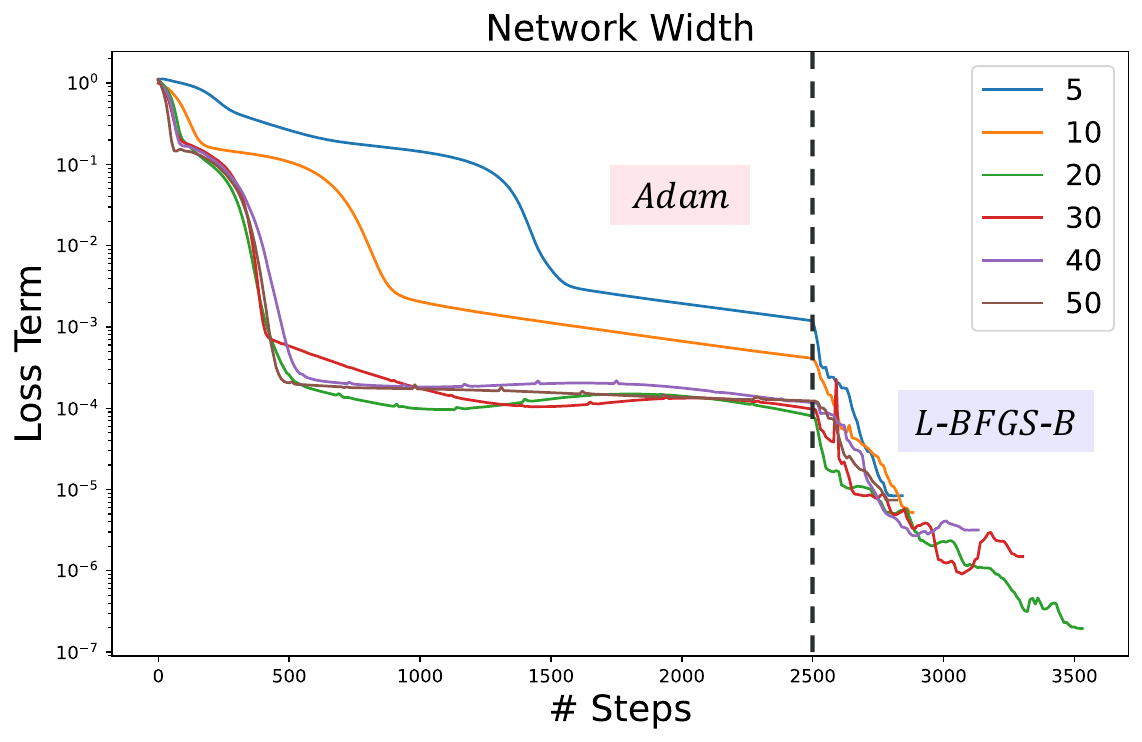}
    \caption{There is no significant improvement in accuracy for anything more than twenty nodes wide.}
    \label{fig:network_width}
\end{figure}

Next, we vary the number of training steps from the Adam optimizer before L-BFGS-B is applied. PINNs are typically trained with two optimizers in succession: Adam and then L-BFGS-B.~\cite{markidis2021old, liu1989limited} Since L-BFGS-B is a second-order optimization method, it converges to minima more accurately. However, L-BFGS-B can rapidly converge and get stuck to local minima. This is why we use Adam before L-BFGS-B to avoid local minima. In Fig.~\ref{fig:adam_bfgs_hpt}, our runs show that Adam is helpful for robust training, since a low number of Adam training steps results in high variation depending on the initialization.

Next, we vary the size of the network, first the depth of the network (i.e. the number of hidden layers) and then the width of the network (i.e. the number of nodes for each hidden layer). We find that anything more than 3 hidden layers does not show a significant increase in the performance of the model as shown in Fig. ~\ref{fig:depth_hpt}. For the width, anything more than 20 nodes does not show a significant increase in the performance as shown in Fig. ~\ref{fig:network_width}. This is probably due to the smooth and relatively well-behaved nature of the Grad-Shafranov solutions with linear profiles.

\bibliography{PINNs}

\providecommand{\noopsort}[1]{}\providecommand{\singleletter}[1]{#1}%
\begin{thebibliography}{40}%
\makeatletter
\providecommand \@ifxundefined [1]{%
 \@ifx{#1\undefined}
}%
\providecommand \@ifnum [1]{%
 \ifnum #1\expandafter \@firstoftwo
 \else \expandafter \@secondoftwo
 \fi
}%
\providecommand \@ifx [1]{%
 \ifx #1\expandafter \@firstoftwo
 \else \expandafter \@secondoftwo
 \fi
}%
\providecommand \natexlab [1]{#1}%
\providecommand \enquote  [1]{``#1''}%
\providecommand \bibnamefont  [1]{#1}%
\providecommand \bibfnamefont [1]{#1}%
\providecommand \citenamefont [1]{#1}%
\providecommand \href@noop [0]{\@secondoftwo}%
\providecommand \href [0]{\begingroup \@sanitize@url \@href}%
\providecommand \@href[1]{\@@startlink{#1}\@@href}%
\providecommand \@@href[1]{\endgroup#1\@@endlink}%
\providecommand \@sanitize@url [0]{\catcode `\\12\catcode `\$12\catcode `\&12\catcode `\#12\catcode `\^12\catcode `\_12\catcode `\%12\relax}%
\providecommand \@@startlink[1]{}%
\providecommand \@@endlink[0]{}%
\providecommand \url  [0]{\begingroup\@sanitize@url \@url }%
\providecommand \@url [1]{\endgroup\@href {#1}{\urlprefix }}%
\providecommand \urlprefix  [0]{URL }%
\providecommand \Eprint [0]{\href }%
\providecommand \doibase [0]{http://dx.doi.org/}%
\providecommand \selectlanguage [0]{\@gobble}%
\providecommand \bibinfo  [0]{\@secondoftwo}%
\providecommand \bibfield  [0]{\@secondoftwo}%
\providecommand \translation [1]{[#1]}%
\providecommand \BibitemOpen [0]{}%
\providecommand \bibitemStop [0]{}%
\providecommand \bibitemNoStop [0]{.\EOS\space}%
\providecommand \EOS [0]{\spacefactor3000\relax}%
\providecommand \BibitemShut  [1]{\csname bibitem#1\endcsname}%
\let\auto@bib@innerbib\@empty
\bibitem [{\citenamefont {Grad}\ and\ \citenamefont {Rubin}(1958)}]{grad1958hydromagnetic}%
  \BibitemOpen
  \bibfield  {author} {\bibinfo {author} {\bibfnamefont {H.}~\bibnamefont {Grad}}\ and\ \bibinfo {author} {\bibfnamefont {H.}~\bibnamefont {Rubin}},\ }\bibfield  {title} {\enquote {\bibinfo {title} {Hydromagnetic equilibria and force-free fields},}\ }\href@noop {} {\bibfield  {journal} {\bibinfo  {journal} {Journal of Nuclear Energy (1954)}\ }\textbf {\bibinfo {volume} {7}},\ \bibinfo {pages} {284--285} (\bibinfo {year} {1958})}\BibitemShut {NoStop}%
\bibitem [{\citenamefont {Shafranov}(1958)}]{shafranov1958magnetohydrodynamical}%
  \BibitemOpen
  \bibfield  {author} {\bibinfo {author} {\bibfnamefont {V.}~\bibnamefont {Shafranov}},\ }\bibfield  {title} {\enquote {\bibinfo {title} {On magnetohydrodynamical equilibrium configurations},}\ }\href@noop {} {\bibfield  {journal} {\bibinfo  {journal} {Soviet Physics JETP}\ }\textbf {\bibinfo {volume} {6}},\ \bibinfo {pages} {1013} (\bibinfo {year} {1958})}\BibitemShut {NoStop}%
\bibitem [{\citenamefont {Lao}\ \emph {et~al.}(2005)\citenamefont {Lao}, \citenamefont {John}, \citenamefont {Peng}, \citenamefont {Ferron}, \citenamefont {Strait}, \citenamefont {Taylor}, \citenamefont {Meyer}, \citenamefont {Zhang},\ and\ \citenamefont {You}}]{lao2005mhd}%
  \BibitemOpen
  \bibfield  {author} {\bibinfo {author} {\bibfnamefont {L.}~\bibnamefont {Lao}}, \bibinfo {author} {\bibfnamefont {H.~S.}\ \bibnamefont {John}}, \bibinfo {author} {\bibfnamefont {Q.}~\bibnamefont {Peng}}, \bibinfo {author} {\bibfnamefont {J.}~\bibnamefont {Ferron}}, \bibinfo {author} {\bibfnamefont {E.}~\bibnamefont {Strait}}, \bibinfo {author} {\bibfnamefont {T.}~\bibnamefont {Taylor}}, \bibinfo {author} {\bibfnamefont {W.}~\bibnamefont {Meyer}}, \bibinfo {author} {\bibfnamefont {C.}~\bibnamefont {Zhang}}, \ and\ \bibinfo {author} {\bibfnamefont {K.}~\bibnamefont {You}},\ }\bibfield  {title} {\enquote {\bibinfo {title} {{MHD} equilibrium reconstruction in the {DIII-D} tokamak},}\ }\href@noop {} {\bibfield  {journal} {\bibinfo  {journal} {Fusion science and technology}\ }\textbf {\bibinfo {volume} {48}},\ \bibinfo {pages} {968--977} (\bibinfo {year} {2005})}\BibitemShut {NoStop}%
\bibitem [{\citenamefont {Raissi}, \citenamefont {Perdikaris},\ and\ \citenamefont {Karniadakis}(2017{\natexlab{a}})}]{raissi2017physics1}%
  \BibitemOpen
  \bibfield  {author} {\bibinfo {author} {\bibfnamefont {M.}~\bibnamefont {Raissi}}, \bibinfo {author} {\bibfnamefont {P.}~\bibnamefont {Perdikaris}}, \ and\ \bibinfo {author} {\bibfnamefont {G.~E.}\ \bibnamefont {Karniadakis}},\ }\bibfield  {title} {\enquote {\bibinfo {title} {Physics informed deep learning (part i): Data-driven solutions of nonlinear partial differential equations},}\ }\href@noop {} {\bibfield  {journal} {\bibinfo  {journal} {arXiv preprint arXiv:1711.10561}\ } (\bibinfo {year} {2017}{\natexlab{a}})}\BibitemShut {NoStop}%
\bibitem [{\citenamefont {Raissi}, \citenamefont {Perdikaris},\ and\ \citenamefont {Karniadakis}(2017{\natexlab{b}})}]{raissi2017physics2}%
  \BibitemOpen
  \bibfield  {author} {\bibinfo {author} {\bibfnamefont {M.}~\bibnamefont {Raissi}}, \bibinfo {author} {\bibfnamefont {P.}~\bibnamefont {Perdikaris}}, \ and\ \bibinfo {author} {\bibfnamefont {G.~E.}\ \bibnamefont {Karniadakis}},\ }\bibfield  {title} {\enquote {\bibinfo {title} {Physics informed deep learning (part ii): Data-driven discovery of nonlinear partial differential equations},}\ }\href@noop {} {\bibfield  {journal} {\bibinfo  {journal} {arXiv preprint arXiv:1711.10566}\ } (\bibinfo {year} {2017}{\natexlab{b}})}\BibitemShut {NoStop}%
\bibitem [{\citenamefont {Blechschmidt}\ and\ \citenamefont {Ernst}(2021)}]{blechschmidt2021three}%
  \BibitemOpen
  \bibfield  {author} {\bibinfo {author} {\bibfnamefont {J.}~\bibnamefont {Blechschmidt}}\ and\ \bibinfo {author} {\bibfnamefont {O.~G.}\ \bibnamefont {Ernst}},\ }\bibfield  {title} {\enquote {\bibinfo {title} {Three ways to solve partial differential equations with neural networks—a review},}\ }\href@noop {} {\bibfield  {journal} {\bibinfo  {journal} {GAMM-Mitteilungen}\ }\textbf {\bibinfo {volume} {44}},\ \bibinfo {pages} {e202100006} (\bibinfo {year} {2021})}\BibitemShut {NoStop}%
\bibitem [{\citenamefont {Karniadakis}\ \emph {et~al.}(2021)\citenamefont {Karniadakis}, \citenamefont {Kevrekidis}, \citenamefont {Lu}, \citenamefont {Perdikaris}, \citenamefont {Wang},\ and\ \citenamefont {Yang}}]{karniadakis2021physics}%
  \BibitemOpen
  \bibfield  {author} {\bibinfo {author} {\bibfnamefont {G.~E.}\ \bibnamefont {Karniadakis}}, \bibinfo {author} {\bibfnamefont {I.~G.}\ \bibnamefont {Kevrekidis}}, \bibinfo {author} {\bibfnamefont {L.}~\bibnamefont {Lu}}, \bibinfo {author} {\bibfnamefont {P.}~\bibnamefont {Perdikaris}}, \bibinfo {author} {\bibfnamefont {S.}~\bibnamefont {Wang}}, \ and\ \bibinfo {author} {\bibfnamefont {L.}~\bibnamefont {Yang}},\ }\bibfield  {title} {\enquote {\bibinfo {title} {Physics-informed machine learning},}\ }\href@noop {} {\bibfield  {journal} {\bibinfo  {journal} {Nature Reviews Physics}\ }\textbf {\bibinfo {volume} {3}},\ \bibinfo {pages} {422--440} (\bibinfo {year} {2021})}\BibitemShut {NoStop}%
\bibitem [{\citenamefont {Shukla}\ \emph {et~al.}(2023)\citenamefont {Shukla}, \citenamefont {Oommen}, \citenamefont {Peyvan}, \citenamefont {Penwarden}, \citenamefont {Bravo}, \citenamefont {Ghoshal}, \citenamefont {Kirby},\ and\ \citenamefont {Karniadakis}}]{shukla2023deep}%
  \BibitemOpen
  \bibfield  {author} {\bibinfo {author} {\bibfnamefont {K.}~\bibnamefont {Shukla}}, \bibinfo {author} {\bibfnamefont {V.}~\bibnamefont {Oommen}}, \bibinfo {author} {\bibfnamefont {A.}~\bibnamefont {Peyvan}}, \bibinfo {author} {\bibfnamefont {M.}~\bibnamefont {Penwarden}}, \bibinfo {author} {\bibfnamefont {L.}~\bibnamefont {Bravo}}, \bibinfo {author} {\bibfnamefont {A.}~\bibnamefont {Ghoshal}}, \bibinfo {author} {\bibfnamefont {R.~M.}\ \bibnamefont {Kirby}}, \ and\ \bibinfo {author} {\bibfnamefont {G.~E.}\ \bibnamefont {Karniadakis}},\ }\bibfield  {title} {\enquote {\bibinfo {title} {Deep neural operators can serve as accurate surrogates for shape optimization: a case study for airfoils},}\ }\href@noop {} {\bibfield  {journal} {\bibinfo  {journal} {arXiv preprint arXiv:2302.00807}\ } (\bibinfo {year} {2023})}\BibitemShut {NoStop}%
\bibitem [{\citenamefont {Rossi}, \citenamefont {Gelfusa},\ and\ \citenamefont {Murari}(2023)}]{rossi2023potential}%
  \BibitemOpen
  \bibfield  {author} {\bibinfo {author} {\bibfnamefont {R.}~\bibnamefont {Rossi}}, \bibinfo {author} {\bibfnamefont {M.}~\bibnamefont {Gelfusa}}, \ and\ \bibinfo {author} {\bibfnamefont {A.}~\bibnamefont {Murari}},\ }\bibfield  {title} {\enquote {\bibinfo {title} {On the potential of physics-informed neural networks to solve inverse problems in tokamaks},}\ }\href@noop {} {\bibfield  {journal} {\bibinfo  {journal} {Nuclear Fusion}\ } (\bibinfo {year} {2023})}\BibitemShut {NoStop}%
\bibitem [{\citenamefont {Cai}\ \emph {et~al.}(2021)\citenamefont {Cai}, \citenamefont {Wang}, \citenamefont {Wang}, \citenamefont {Perdikaris},\ and\ \citenamefont {Karniadakis}}]{cai2021physics}%
  \BibitemOpen
  \bibfield  {author} {\bibinfo {author} {\bibfnamefont {S.}~\bibnamefont {Cai}}, \bibinfo {author} {\bibfnamefont {Z.}~\bibnamefont {Wang}}, \bibinfo {author} {\bibfnamefont {S.}~\bibnamefont {Wang}}, \bibinfo {author} {\bibfnamefont {P.}~\bibnamefont {Perdikaris}}, \ and\ \bibinfo {author} {\bibfnamefont {G.~E.}\ \bibnamefont {Karniadakis}},\ }\bibfield  {title} {\enquote {\bibinfo {title} {Physics-informed neural networks for heat transfer problems},}\ }\href@noop {} {\bibfield  {journal} {\bibinfo  {journal} {Journal of Heat Transfer}\ }\textbf {\bibinfo {volume} {143}} (\bibinfo {year} {2021})}\BibitemShut {NoStop}%
\bibitem [{\citenamefont {van Milligen}, \citenamefont {Tribaldos},\ and\ \citenamefont {Jim{\'e}nez}(1995)}]{van1995neural}%
  \BibitemOpen
  \bibfield  {author} {\bibinfo {author} {\bibfnamefont {B.~P.}\ \bibnamefont {van Milligen}}, \bibinfo {author} {\bibfnamefont {V.}~\bibnamefont {Tribaldos}}, \ and\ \bibinfo {author} {\bibfnamefont {J.}~\bibnamefont {Jim{\'e}nez}},\ }\bibfield  {title} {\enquote {\bibinfo {title} {Neural network differential equation and plasma equilibrium solver},}\ }\href@noop {} {\bibfield  {journal} {\bibinfo  {journal} {Physical review letters}\ }\textbf {\bibinfo {volume} {75}},\ \bibinfo {pages} {3594} (\bibinfo {year} {1995})}\BibitemShut {NoStop}%
\bibitem [{\citenamefont {Joung}\ \emph {et~al.}(2019)\citenamefont {Joung}, \citenamefont {Kim}, \citenamefont {Kwak}, \citenamefont {Bak}, \citenamefont {Lee}, \citenamefont {Han}, \citenamefont {Kim}, \citenamefont {Lee}, \citenamefont {Kwon},\ and\ \citenamefont {Ghim}}]{joung2019deep}%
  \BibitemOpen
  \bibfield  {author} {\bibinfo {author} {\bibfnamefont {S.}~\bibnamefont {Joung}}, \bibinfo {author} {\bibfnamefont {J.}~\bibnamefont {Kim}}, \bibinfo {author} {\bibfnamefont {S.}~\bibnamefont {Kwak}}, \bibinfo {author} {\bibfnamefont {J.}~\bibnamefont {Bak}}, \bibinfo {author} {\bibfnamefont {S.}~\bibnamefont {Lee}}, \bibinfo {author} {\bibfnamefont {H.}~\bibnamefont {Han}}, \bibinfo {author} {\bibfnamefont {H.}~\bibnamefont {Kim}}, \bibinfo {author} {\bibfnamefont {G.}~\bibnamefont {Lee}}, \bibinfo {author} {\bibfnamefont {D.}~\bibnamefont {Kwon}}, \ and\ \bibinfo {author} {\bibfnamefont {Y.-C.}\ \bibnamefont {Ghim}},\ }\bibfield  {title} {\enquote {\bibinfo {title} {Deep neural network {Grad--Shafranov} solver constrained with measured magnetic signals},}\ }\href@noop {} {\bibfield  {journal} {\bibinfo  {journal} {Nuclear Fusion}\ }\textbf {\bibinfo {volume} {60}},\ \bibinfo {pages} {016034} (\bibinfo {year} {2019})}\BibitemShut {NoStop}%
\bibitem [{\citenamefont {Merlo}\ \emph {et~al.}(2021)\citenamefont {Merlo}, \citenamefont {B{\"o}ckenhoff}, \citenamefont {Schilling}, \citenamefont {H{\"o}fel}, \citenamefont {Kwak}, \citenamefont {Svensson}, \citenamefont {Pavone}, \citenamefont {Lazerson}, \citenamefont {Pedersen} \emph {et~al.}}]{merlo2021proof}%
  \BibitemOpen
  \bibfield  {author} {\bibinfo {author} {\bibfnamefont {A.}~\bibnamefont {Merlo}}, \bibinfo {author} {\bibfnamefont {D.}~\bibnamefont {B{\"o}ckenhoff}}, \bibinfo {author} {\bibfnamefont {J.}~\bibnamefont {Schilling}}, \bibinfo {author} {\bibfnamefont {U.}~\bibnamefont {H{\"o}fel}}, \bibinfo {author} {\bibfnamefont {S.}~\bibnamefont {Kwak}}, \bibinfo {author} {\bibfnamefont {J.}~\bibnamefont {Svensson}}, \bibinfo {author} {\bibfnamefont {A.}~\bibnamefont {Pavone}}, \bibinfo {author} {\bibfnamefont {S.~A.}\ \bibnamefont {Lazerson}}, \bibinfo {author} {\bibfnamefont {T.~S.}\ \bibnamefont {Pedersen}},  \emph {et~al.},\ }\bibfield  {title} {\enquote {\bibinfo {title} {Proof of concept of a fast surrogate model of the {VMEC} code via neural networks in {Wendelstein 7-X} scenarios},}\ }\href@noop {} {\bibfield  {journal} {\bibinfo  {journal} {Nuclear Fusion}\ }\textbf {\bibinfo {volume} {61}},\ \bibinfo {pages} {096039} (\bibinfo {year} {2021})}\BibitemShut {NoStop}%
\bibitem [{\citenamefont {Wai}, \citenamefont {Boyer},\ and\ \citenamefont {Kolemen}(2022)}]{wai2022neural}%
  \BibitemOpen
  \bibfield  {author} {\bibinfo {author} {\bibfnamefont {J.}~\bibnamefont {Wai}}, \bibinfo {author} {\bibfnamefont {M.}~\bibnamefont {Boyer}}, \ and\ \bibinfo {author} {\bibfnamefont {E.}~\bibnamefont {Kolemen}},\ }\bibfield  {title} {\enquote {\bibinfo {title} {Neural net modeling of equilibria in {NSTX-U}},}\ }\href@noop {} {\bibfield  {journal} {\bibinfo  {journal} {arXiv preprint arXiv:2202.13915}\ } (\bibinfo {year} {2022})}\BibitemShut {NoStop}%
\bibitem [{\citenamefont {Liu}\ \emph {et~al.}(2022)\citenamefont {Liu}, \citenamefont {Akcay}, \citenamefont {Lao},\ and\ \citenamefont {Sun}}]{liu2022surrogate}%
  \BibitemOpen
  \bibfield  {author} {\bibinfo {author} {\bibfnamefont {Y.}~\bibnamefont {Liu}}, \bibinfo {author} {\bibfnamefont {C.}~\bibnamefont {Akcay}}, \bibinfo {author} {\bibfnamefont {L.~L.}\ \bibnamefont {Lao}}, \ and\ \bibinfo {author} {\bibfnamefont {X.}~\bibnamefont {Sun}},\ }\bibfield  {title} {\enquote {\bibinfo {title} {Surrogate models for plasma displacement and current in {3-D} perturbed magnetohydrodynamic equilibria in tokamaks},}\ }\href@noop {} {\bibfield  {journal} {\bibinfo  {journal} {Nuclear Fusion}\ } (\bibinfo {year} {2022})}\BibitemShut {NoStop}%
\bibitem [{\citenamefont {Merlo}\ \emph {et~al.}(2023)\citenamefont {Merlo}, \citenamefont {B{\"o}ckenhoff}, \citenamefont {Schilling}, \citenamefont {Lazerson}, \citenamefont {Pedersen} \emph {et~al.}}]{merlo2023physics}%
  \BibitemOpen
  \bibfield  {author} {\bibinfo {author} {\bibfnamefont {A.}~\bibnamefont {Merlo}}, \bibinfo {author} {\bibfnamefont {D.}~\bibnamefont {B{\"o}ckenhoff}}, \bibinfo {author} {\bibfnamefont {J.}~\bibnamefont {Schilling}}, \bibinfo {author} {\bibfnamefont {S.~A.}\ \bibnamefont {Lazerson}}, \bibinfo {author} {\bibfnamefont {T.~S.}\ \bibnamefont {Pedersen}},  \emph {et~al.},\ }\bibfield  {title} {\enquote {\bibinfo {title} {Physics-regularized neural network of the ideal-mhd solution operator in wendelstein 7-x configurations},}\ }\href@noop {} {\bibfield  {journal} {\bibinfo  {journal} {Nuclear Fusion}\ }\textbf {\bibinfo {volume} {63}},\ \bibinfo {pages} {066020} (\bibinfo {year} {2023})}\BibitemShut {NoStop}%
\bibitem [{\citenamefont {Kaltsas}\ and\ \citenamefont {Throumoulopoulos}(2022)}]{kaltsas2022neural}%
  \BibitemOpen
  \bibfield  {author} {\bibinfo {author} {\bibfnamefont {D.}~\bibnamefont {Kaltsas}}\ and\ \bibinfo {author} {\bibfnamefont {G.}~\bibnamefont {Throumoulopoulos}},\ }\bibfield  {title} {\enquote {\bibinfo {title} {Neural network tokamak equilibria with incompressible flows},}\ }\href@noop {} {\bibfield  {journal} {\bibinfo  {journal} {Physics of Plasmas}\ }\textbf {\bibinfo {volume} {29}} (\bibinfo {year} {2022})}\BibitemShut {NoStop}%
\bibitem [{\citenamefont {Cerfon}\ and\ \citenamefont {Freidberg}(2010)}]{cerfon2010one}%
  \BibitemOpen
  \bibfield  {author} {\bibinfo {author} {\bibfnamefont {A.~J.}\ \bibnamefont {Cerfon}}\ and\ \bibinfo {author} {\bibfnamefont {J.~P.}\ \bibnamefont {Freidberg}},\ }\bibfield  {title} {\enquote {\bibinfo {title} {“one size fits all” analytic solutions to the {G}rad--{S}hafranov equation},}\ }\href@noop {} {\bibfield  {journal} {\bibinfo  {journal} {Physics of Plasmas}\ }\textbf {\bibinfo {volume} {17}},\ \bibinfo {pages} {032502} (\bibinfo {year} {2010})}\BibitemShut {NoStop}%
\bibitem [{\citenamefont {Lu}\ \emph {et~al.}(2021{\natexlab{a}})\citenamefont {Lu}, \citenamefont {Meng}, \citenamefont {Mao},\ and\ \citenamefont {Karniadakis}}]{lu2021deepxde}%
  \BibitemOpen
  \bibfield  {author} {\bibinfo {author} {\bibfnamefont {L.}~\bibnamefont {Lu}}, \bibinfo {author} {\bibfnamefont {X.}~\bibnamefont {Meng}}, \bibinfo {author} {\bibfnamefont {Z.}~\bibnamefont {Mao}}, \ and\ \bibinfo {author} {\bibfnamefont {G.~E.}\ \bibnamefont {Karniadakis}},\ }\bibfield  {title} {\enquote {\bibinfo {title} {Deepxde: A deep learning library for solving differential equations},}\ }\href@noop {} {\bibfield  {journal} {\bibinfo  {journal} {SIAM review}\ }\textbf {\bibinfo {volume} {63}},\ \bibinfo {pages} {208--228} (\bibinfo {year} {2021}{\natexlab{a}})}\BibitemShut {NoStop}%
\bibitem [{\citenamefont {Freidberg}(2014)}]{freidberg2014ideal}%
  \BibitemOpen
  \bibfield  {author} {\bibinfo {author} {\bibfnamefont {J.~P.}\ \bibnamefont {Freidberg}},\ }\href@noop {} {\emph {\bibinfo {title} {ideal MHD}}}\ (\bibinfo  {publisher} {Cambridge University Press},\ \bibinfo {year} {2014})\BibitemShut {NoStop}%
\bibitem [{\citenamefont {Hazeltine}\ and\ \citenamefont {Meiss}(2003)}]{hazeltine2003plasma}%
  \BibitemOpen
  \bibfield  {author} {\bibinfo {author} {\bibfnamefont {R.~D.}\ \bibnamefont {Hazeltine}}\ and\ \bibinfo {author} {\bibfnamefont {J.~D.}\ \bibnamefont {Meiss}},\ }\href@noop {} {\emph {\bibinfo {title} {Plasma confinement}}}\ (\bibinfo  {publisher} {Courier Corporation},\ \bibinfo {year} {2003})\BibitemShut {NoStop}%
\bibitem [{\citenamefont {Solov’ev}(1968)}]{solov1968theory}%
  \BibitemOpen
  \bibfield  {author} {\bibinfo {author} {\bibfnamefont {L.}~\bibnamefont {Solov’ev}},\ }\bibfield  {title} {\enquote {\bibinfo {title} {The theory of hydromagnetic stability of toroidal plasma configurations},}\ }\href@noop {} {\bibfield  {journal} {\bibinfo  {journal} {Sov. Phys. JETP}\ }\textbf {\bibinfo {volume} {26}},\ \bibinfo {pages} {400--407} (\bibinfo {year} {1968})}\BibitemShut {NoStop}%
\bibitem [{\citenamefont {Zheng}\ \emph {et~al.}(2018)\citenamefont {Zheng}, \citenamefont {Hu}, \citenamefont {Zhang}, \citenamefont {Chen}, \citenamefont {Zhao}, \citenamefont {Wang}, \citenamefont {Shi}, \citenamefont {Zhang}, \citenamefont {Zhang}, \citenamefont {Zhou}, \citenamefont {Wei},\ and\ \citenamefont {and}}]{Zheng_2018}%
  \BibitemOpen
  \bibfield  {author} {\bibinfo {author} {\bibfnamefont {W.}~\bibnamefont {Zheng}}, \bibinfo {author} {\bibfnamefont {F.}~\bibnamefont {Hu}}, \bibinfo {author} {\bibfnamefont {M.}~\bibnamefont {Zhang}}, \bibinfo {author} {\bibfnamefont {Z.}~\bibnamefont {Chen}}, \bibinfo {author} {\bibfnamefont {X.}~\bibnamefont {Zhao}}, \bibinfo {author} {\bibfnamefont {X.}~\bibnamefont {Wang}}, \bibinfo {author} {\bibfnamefont {P.}~\bibnamefont {Shi}}, \bibinfo {author} {\bibfnamefont {X.}~\bibnamefont {Zhang}}, \bibinfo {author} {\bibfnamefont {X.}~\bibnamefont {Zhang}}, \bibinfo {author} {\bibfnamefont {Y.}~\bibnamefont {Zhou}}, \bibinfo {author} {\bibfnamefont {Y.}~\bibnamefont {Wei}}, \ and\ \bibinfo {author} {\bibfnamefont {Y.~P.}\ \bibnamefont {and}},\ }\bibfield  {title} {\enquote {\bibinfo {title} {Hybrid neural network for density limit disruption prediction and avoidance on j-{TEXT} tokamak},}\ }\href {\doibase 10.1088/1741-4326/aaad17} {\bibfield  {journal} {\bibinfo  {journal} {Nuclear Fusion}\ }\textbf
  {\bibinfo {volume} {58}},\ \bibinfo {pages} {056016} (\bibinfo {year} {2018})}\BibitemShut {NoStop}%
\bibitem [{\citenamefont {Hornik}, \citenamefont {Stinchcombe},\ and\ \citenamefont {White}(1989)}]{hornik1989multilayer}%
  \BibitemOpen
  \bibfield  {author} {\bibinfo {author} {\bibfnamefont {K.}~\bibnamefont {Hornik}}, \bibinfo {author} {\bibfnamefont {M.}~\bibnamefont {Stinchcombe}}, \ and\ \bibinfo {author} {\bibfnamefont {H.}~\bibnamefont {White}},\ }\bibfield  {title} {\enquote {\bibinfo {title} {Multilayer feedforward networks are universal approximators},}\ }\href@noop {} {\bibfield  {journal} {\bibinfo  {journal} {Neural Netw.}\ }\textbf {\bibinfo {volume} {2}},\ \bibinfo {pages} {359--366} (\bibinfo {year} {1989})}\BibitemShut {NoStop}%
\bibitem [{\citenamefont {Raissi}, \citenamefont {Perdikaris},\ and\ \citenamefont {Karniadakis}(2019)}]{raissi2019physics}%
  \BibitemOpen
  \bibfield  {author} {\bibinfo {author} {\bibfnamefont {M.}~\bibnamefont {Raissi}}, \bibinfo {author} {\bibfnamefont {P.}~\bibnamefont {Perdikaris}}, \ and\ \bibinfo {author} {\bibfnamefont {G.~E.}\ \bibnamefont {Karniadakis}},\ }\bibfield  {title} {\enquote {\bibinfo {title} {Physics-informed neural networks: A deep learning framework for solving forward and inverse problems involving nonlinear partial differential equations},}\ }\href@noop {} {\bibfield  {journal} {\bibinfo  {journal} {Journal of Computational Physics}\ }\textbf {\bibinfo {volume} {378}},\ \bibinfo {pages} {686--707} (\bibinfo {year} {2019})}\BibitemShut {NoStop}%
\bibitem [{\citenamefont {Sun}\ \emph {et~al.}(2020)\citenamefont {Sun}, \citenamefont {Gao}, \citenamefont {Pan},\ and\ \citenamefont {Wang}}]{sun2020surrogate}%
  \BibitemOpen
  \bibfield  {author} {\bibinfo {author} {\bibfnamefont {L.}~\bibnamefont {Sun}}, \bibinfo {author} {\bibfnamefont {H.}~\bibnamefont {Gao}}, \bibinfo {author} {\bibfnamefont {S.}~\bibnamefont {Pan}}, \ and\ \bibinfo {author} {\bibfnamefont {J.-X.}\ \bibnamefont {Wang}},\ }\bibfield  {title} {\enquote {\bibinfo {title} {Surrogate modeling for fluid flows based on physics-constrained deep learning without simulation data},}\ }\href@noop {} {\bibfield  {journal} {\bibinfo  {journal} {Computer Methods in Applied Mechanics and Engineering}\ }\textbf {\bibinfo {volume} {361}},\ \bibinfo {pages} {112732} (\bibinfo {year} {2020})}\BibitemShut {NoStop}%
\bibitem [{\citenamefont {Lu}\ \emph {et~al.}(2021{\natexlab{b}})\citenamefont {Lu}, \citenamefont {Pestourie}, \citenamefont {Yao}, \citenamefont {Wang}, \citenamefont {Verdugo},\ and\ \citenamefont {Johnson}}]{lu2021physics}%
  \BibitemOpen
  \bibfield  {author} {\bibinfo {author} {\bibfnamefont {L.}~\bibnamefont {Lu}}, \bibinfo {author} {\bibfnamefont {R.}~\bibnamefont {Pestourie}}, \bibinfo {author} {\bibfnamefont {W.}~\bibnamefont {Yao}}, \bibinfo {author} {\bibfnamefont {Z.}~\bibnamefont {Wang}}, \bibinfo {author} {\bibfnamefont {F.}~\bibnamefont {Verdugo}}, \ and\ \bibinfo {author} {\bibfnamefont {S.~G.}\ \bibnamefont {Johnson}},\ }\bibfield  {title} {\enquote {\bibinfo {title} {Physics-informed neural networks with hard constraints for inverse design},}\ }\href@noop {} {\bibfield  {journal} {\bibinfo  {journal} {arXiv preprint arXiv:2102.04626}\ } (\bibinfo {year} {2021}{\natexlab{b}})}\BibitemShut {NoStop}%
\bibitem [{\citenamefont {Kingma}\ and\ \citenamefont {Ba}(2014)}]{kingma2014adam}%
  \BibitemOpen
  \bibfield  {author} {\bibinfo {author} {\bibfnamefont {D.~P.}\ \bibnamefont {Kingma}}\ and\ \bibinfo {author} {\bibfnamefont {J.}~\bibnamefont {Ba}},\ }\bibfield  {title} {\enquote {\bibinfo {title} {Adam: A method for stochastic optimization},}\ }\href@noop {} {\bibfield  {journal} {\bibinfo  {journal} {arXiv preprint arXiv:1412.6980}\ } (\bibinfo {year} {2014})}\BibitemShut {NoStop}%
\bibitem [{\citenamefont {Zhu}\ \emph {et~al.}(1997)\citenamefont {Zhu}, \citenamefont {Byrd}, \citenamefont {Lu},\ and\ \citenamefont {Nocedal}}]{zhu1997algorithm}%
  \BibitemOpen
  \bibfield  {author} {\bibinfo {author} {\bibfnamefont {C.}~\bibnamefont {Zhu}}, \bibinfo {author} {\bibfnamefont {R.~H.}\ \bibnamefont {Byrd}}, \bibinfo {author} {\bibfnamefont {P.}~\bibnamefont {Lu}}, \ and\ \bibinfo {author} {\bibfnamefont {J.}~\bibnamefont {Nocedal}},\ }\bibfield  {title} {\enquote {\bibinfo {title} {Algorithm 778: L-bfgs-b: Fortran subroutines for large-scale bound-constrained optimization},}\ }\href@noop {} {\bibfield  {journal} {\bibinfo  {journal} {ACM Transactions on mathematical software (TOMS)}\ }\textbf {\bibinfo {volume} {23}},\ \bibinfo {pages} {550--560} (\bibinfo {year} {1997})}\BibitemShut {NoStop}%
\bibitem [{\citenamefont {Markidis}(2021)}]{markidis2021old}%
  \BibitemOpen
  \bibfield  {author} {\bibinfo {author} {\bibfnamefont {S.}~\bibnamefont {Markidis}},\ }\bibfield  {title} {\enquote {\bibinfo {title} {The old and the new: Can physics-informed deep-learning replace traditional linear solvers?}}\ }\href@noop {} {\bibfield  {journal} {\bibinfo  {journal} {Frontiers in big Data}\ }\textbf {\bibinfo {volume} {4}},\ \bibinfo {pages} {669097} (\bibinfo {year} {2021})}\BibitemShut {NoStop}%
\bibitem [{\citenamefont {Tang}, \citenamefont {Wan},\ and\ \citenamefont {Yang}(2023)}]{tang2023pinns}%
  \BibitemOpen
  \bibfield  {author} {\bibinfo {author} {\bibfnamefont {K.}~\bibnamefont {Tang}}, \bibinfo {author} {\bibfnamefont {X.}~\bibnamefont {Wan}}, \ and\ \bibinfo {author} {\bibfnamefont {C.}~\bibnamefont {Yang}},\ }\bibfield  {title} {\enquote {\bibinfo {title} {Das-pinns: A deep adaptive sampling method for solving high-dimensional partial differential equations},}\ }\href@noop {} {\bibfield  {journal} {\bibinfo  {journal} {Journal of Computational Physics}\ }\textbf {\bibinfo {volume} {476}},\ \bibinfo {pages} {111868} (\bibinfo {year} {2023})}\BibitemShut {NoStop}%
\bibitem [{\citenamefont {Wu}\ \emph {et~al.}(2023)\citenamefont {Wu}, \citenamefont {Zhu}, \citenamefont {Tan}, \citenamefont {Kartha},\ and\ \citenamefont {Lu}}]{wu2023comprehensive}%
  \BibitemOpen
  \bibfield  {author} {\bibinfo {author} {\bibfnamefont {C.}~\bibnamefont {Wu}}, \bibinfo {author} {\bibfnamefont {M.}~\bibnamefont {Zhu}}, \bibinfo {author} {\bibfnamefont {Q.}~\bibnamefont {Tan}}, \bibinfo {author} {\bibfnamefont {Y.}~\bibnamefont {Kartha}}, \ and\ \bibinfo {author} {\bibfnamefont {L.}~\bibnamefont {Lu}},\ }\bibfield  {title} {\enquote {\bibinfo {title} {A comprehensive study of non-adaptive and residual-based adaptive sampling for physics-informed neural networks},}\ }\href@noop {} {\bibfield  {journal} {\bibinfo  {journal} {Computer Methods in Applied Mechanics and Engineering}\ }\textbf {\bibinfo {volume} {403}},\ \bibinfo {pages} {115671} (\bibinfo {year} {2023})}\BibitemShut {NoStop}%
\bibitem [{\citenamefont {Cho}\ \emph {et~al.}(2022)\citenamefont {Cho}, \citenamefont {Nam}, \citenamefont {Yang}, \citenamefont {Yun}, \citenamefont {Hong},\ and\ \citenamefont {Park}}]{cho2022separable}%
  \BibitemOpen
  \bibfield  {author} {\bibinfo {author} {\bibfnamefont {J.}~\bibnamefont {Cho}}, \bibinfo {author} {\bibfnamefont {S.}~\bibnamefont {Nam}}, \bibinfo {author} {\bibfnamefont {H.}~\bibnamefont {Yang}}, \bibinfo {author} {\bibfnamefont {S.-B.}\ \bibnamefont {Yun}}, \bibinfo {author} {\bibfnamefont {Y.}~\bibnamefont {Hong}}, \ and\ \bibinfo {author} {\bibfnamefont {E.}~\bibnamefont {Park}},\ }\bibfield  {title} {\enquote {\bibinfo {title} {Separable pinn: Mitigating the curse of dimensionality in physics-informed neural networks},}\ }\href@noop {} {\bibfield  {journal} {\bibinfo  {journal} {arXiv preprint arXiv:2211.08761}\ } (\bibinfo {year} {2022})}\BibitemShut {NoStop}%
\bibitem [{\citenamefont {LeCun}\ \emph {et~al.}(2002)\citenamefont {LeCun}, \citenamefont {Bottou}, \citenamefont {Orr},\ and\ \citenamefont {M{\"u}ller}}]{lecun2002efficient}%
  \BibitemOpen
  \bibfield  {author} {\bibinfo {author} {\bibfnamefont {Y.}~\bibnamefont {LeCun}}, \bibinfo {author} {\bibfnamefont {L.}~\bibnamefont {Bottou}}, \bibinfo {author} {\bibfnamefont {G.~B.}\ \bibnamefont {Orr}}, \ and\ \bibinfo {author} {\bibfnamefont {K.-R.}\ \bibnamefont {M{\"u}ller}},\ }\bibfield  {title} {\enquote {\bibinfo {title} {Efficient backprop},}\ }in\ \href@noop {} {\emph {\bibinfo {booktitle} {Neural networks: Tricks of the trade}}}\ (\bibinfo  {publisher} {Springer},\ \bibinfo {year} {2002})\ pp.\ \bibinfo {pages} {9--50}\BibitemShut {NoStop}%
\bibitem [{\citenamefont {Wang}, \citenamefont {Teng},\ and\ \citenamefont {Perdikaris}(2021)}]{wang2021understanding}%
  \BibitemOpen
  \bibfield  {author} {\bibinfo {author} {\bibfnamefont {S.}~\bibnamefont {Wang}}, \bibinfo {author} {\bibfnamefont {Y.}~\bibnamefont {Teng}}, \ and\ \bibinfo {author} {\bibfnamefont {P.}~\bibnamefont {Perdikaris}},\ }\bibfield  {title} {\enquote {\bibinfo {title} {Understanding and mitigating gradient flow pathologies in physics-informed neural networks},}\ }\href@noop {} {\bibfield  {journal} {\bibinfo  {journal} {SIAM Journal on Scientific Computing}\ }\textbf {\bibinfo {volume} {43}},\ \bibinfo {pages} {A3055--A3081} (\bibinfo {year} {2021})}\BibitemShut {NoStop}%
\bibitem [{\citenamefont {Nair}\ and\ \citenamefont {Hinton}(2010)}]{nair2010rectified}%
  \BibitemOpen
  \bibfield  {author} {\bibinfo {author} {\bibfnamefont {V.}~\bibnamefont {Nair}}\ and\ \bibinfo {author} {\bibfnamefont {G.~E.}\ \bibnamefont {Hinton}},\ }\bibfield  {title} {\enquote {\bibinfo {title} {Rectified linear units improve restricted boltzmann machines},}\ }in\ \href@noop {} {\emph {\bibinfo {booktitle} {Proceedings of the 27th international conference on machine learning (ICML-10)}}}\ (\bibinfo {year} {2010})\ pp.\ \bibinfo {pages} {807--814}\BibitemShut {NoStop}%
\bibitem [{\citenamefont {Agarap}(2018)}]{agarap2018deep}%
  \BibitemOpen
  \bibfield  {author} {\bibinfo {author} {\bibfnamefont {A.~F.}\ \bibnamefont {Agarap}},\ }\bibfield  {title} {\enquote {\bibinfo {title} {Deep learning using rectified linear units (relu)},}\ }\href@noop {} {\bibfield  {journal} {\bibinfo  {journal} {arXiv preprint arXiv:1803.08375}\ } (\bibinfo {year} {2018})}\BibitemShut {NoStop}%
\bibitem [{\citenamefont {Ramachandran}, \citenamefont {Zoph},\ and\ \citenamefont {Le}(2017)}]{ramachandran2017searching}%
  \BibitemOpen
  \bibfield  {author} {\bibinfo {author} {\bibfnamefont {P.}~\bibnamefont {Ramachandran}}, \bibinfo {author} {\bibfnamefont {B.}~\bibnamefont {Zoph}}, \ and\ \bibinfo {author} {\bibfnamefont {Q.~V.}\ \bibnamefont {Le}},\ }\bibfield  {title} {\enquote {\bibinfo {title} {Searching for activation functions},}\ }\href@noop {} {\bibfield  {journal} {\bibinfo  {journal} {arXiv preprint arXiv:1710.05941}\ } (\bibinfo {year} {2017})}\BibitemShut {NoStop}%
\bibitem [{\citenamefont {Wu}\ \emph {et~al.}(2019)\citenamefont {Wu}, \citenamefont {Liu}, \citenamefont {Bae}, \citenamefont {Chow}, \citenamefont {Iyengar}, \citenamefont {Pu}, \citenamefont {Wei}, \citenamefont {Yu},\ and\ \citenamefont {Zhang}}]{wu2019demystifying}%
  \BibitemOpen
  \bibfield  {author} {\bibinfo {author} {\bibfnamefont {Y.}~\bibnamefont {Wu}}, \bibinfo {author} {\bibfnamefont {L.}~\bibnamefont {Liu}}, \bibinfo {author} {\bibfnamefont {J.}~\bibnamefont {Bae}}, \bibinfo {author} {\bibfnamefont {K.-H.}\ \bibnamefont {Chow}}, \bibinfo {author} {\bibfnamefont {A.}~\bibnamefont {Iyengar}}, \bibinfo {author} {\bibfnamefont {C.}~\bibnamefont {Pu}}, \bibinfo {author} {\bibfnamefont {W.}~\bibnamefont {Wei}}, \bibinfo {author} {\bibfnamefont {L.}~\bibnamefont {Yu}}, \ and\ \bibinfo {author} {\bibfnamefont {Q.}~\bibnamefont {Zhang}},\ }\bibfield  {title} {\enquote {\bibinfo {title} {Demystifying learning rate policies for high accuracy training of deep neural networks},}\ }in\ \href@noop {} {\emph {\bibinfo {booktitle} {2019 IEEE International conference on big data (Big Data)}}}\ (\bibinfo {organization} {IEEE},\ \bibinfo {year} {2019})\ pp.\ \bibinfo {pages} {1971--1980}\BibitemShut {NoStop}%
\bibitem [{\citenamefont {Liu}\ and\ \citenamefont {Nocedal}(1989)}]{liu1989limited}%
  \BibitemOpen
  \bibfield  {author} {\bibinfo {author} {\bibfnamefont {D.~C.}\ \bibnamefont {Liu}}\ and\ \bibinfo {author} {\bibfnamefont {J.}~\bibnamefont {Nocedal}},\ }\bibfield  {title} {\enquote {\bibinfo {title} {On the limited memory bfgs method for large scale optimization},}\ }\href@noop {} {\bibfield  {journal} {\bibinfo  {journal} {Mathematical programming}\ }\textbf {\bibinfo {volume} {45}},\ \bibinfo {pages} {503--528} (\bibinfo {year} {1989})}\BibitemShut {NoStop}%
\end{thebibliography}%
\end{document}